\documentclass{PoS}
\rightline{\parbox{3cm}{\large\rm
MKPH-T-07-18\\
SHEP-0744
}}

\usepackage{epsfig}
\usepackage{rotating}
\usepackage{multirow}
\usepackage{psfrag}
\usepackage{axodraw}

\newcommand{\bi}{\begin{itemize}}
\newcommand{\ei}{\end{itemize}}
\newcommand{\bdot}{\textbullet}
\title{Progress in kaon physics on the lattice}

\ShortTitle{Progress in kaon physics on the lattice}

\author{\speaker{Andreas J\"uttner}\footnote{
New address: 
Institut f\"ur Kernphysik,
Johannes Gutenberg-Universit\"at Mainz,
Johann-Joachim-Becher-Weg 45,
D-55099 Mainz, Germany}\\
        School of Physics and Astronomy\\
	University of Southampton\\
	Southampton, SO17 1BJ\\ UK\\
        E-mail: \email{juettner@kph.uni-mainz.de}}

\abstract{
CKM-unitarity, direct and indirect CP-violation and the $\Delta I=1/2$ 
rule in full lattice QCD are the focus of this talk. To this end I will discuss and compare recent lattice results for leptonic, semi-leptonic and non-leptonic decays of the kaon and neutral kaon mixing and I will motivate current
best estimates
 \begin{center} 
$f_K/f_\pi=1.198(10)$\,,\, 
	$f_+^{K\pi}(0)=0.964(5)$\,\, and \,\,$\hat{B}_K=0.720(39)$ \,.
\end{center}
Moreover 
new theoretical advances that will improve the quality of these 
computations in the future will be discussed.
}

\FullConference{The XXV International Symposium on Lattice Field Theory\\
		 July 30-4 August 2007\\
		 Regensburg, Germany}

\begin{document}

\section{Introduction}\mbox{}\\[-8mm]\indent
This talk reviews recent results and 
developments in the phenomenology of the kaon from the
lattice  focussing on simulations of full QCD
(i.e. with a degenerate pair of dynamical $u$ and $d$ quarks ($N_f=2$) and
a dynamical $s$-quark ($N_f=2+1$)).
Many of the results presented here have been computed using different
fermion discretisations. This has to be seen as a feature since we are entering
a phase where in particular observables describing
$SU(3)$-breaking effects in kaons can be computed on the lattice with 
sub-percent
level precision. If used to  constrain the Standard
Model and also to constrain models that go beyond, consistency of the results
from different discretisations will increase the reliability of 
the lattice output. 
To this end I want to draw the reader's attention to the plenary
talks on recent simulations with domain wall fermions (DWF) \cite{Peter},
overlap fermions \cite{Matsufuru}, twisted mass fermions (tmQCD)~\cite{Urbach} 
and 
clover improved Wilson fermions \cite{Kuramashi} and also the two plenary talks
by Creutz \cite{Creutz} and Kronfeld \cite{Kronfeld}\footnote{See also
Sharpe's plenary talk last year \cite{Sharpe:2006re}.} discussing different 
perspectives on the consistency of the staggered fermion (KS) formulation.

Please be aware of the preliminary status of some of the results presented 
here and that I could not cover all recent results and developments of this 
field. 

This talk is structured as follows: I will start
with a discussion of the status of the first-row-unitarity of the
CKM matrix, particularly focusing on the matrix element
$|V_{us}|$ which can be determined either from leptonic kaon 
and pion decays or from semi-leptonic kaon decays. I will then discuss
indirect $CP$-violation in neutral kaon mixing where the lattice can
contribute in terms of the computation of the bag parameter $\hat B_K$.
Finally, I will discuss direct $CP$-violation and the $\Delta I=1/2$-rule
in hadronic kaon decays, where the 
parameters $|\epsilon^\prime/\epsilon|$ and $\omega$ can be computed 
on the lattice.
While discussing these topics I will also present new theoretical 
and technical developments which will hopefully improve the quality and 
the precision of these calculations.
\section{The determination of $|V_{us}|$}\mbox{}\\[-8mm]\indent
The three entries $V_{ud}$, $V_{us}$ and $V_{ub}$ make up the 
first row of the CKM-matrix \cite{Cabibbo:1963yz,Kobayashi:1973fv} 
and assuming its unitarity,  $|V_{ud}|^2+|V_{us}|^2+|V_{ub}|^2=1$. 
Any deviation from 1 on the r.h.s. 
would be a sign for physics beyond the Standard Model. $|V_{ud}|$ is very 
well known and neglecting $|V_{ub}|$ (since it is very small) 
the precision of $|V_{us}|$ is currently 
limiting the accuracy of the unitarity test.
\subsection{$|V_{us}|$ from leptonic decays}\mbox{}\\[-8mm]\indent
In 2004 Marciano \cite{Marciano:2004uf} 
first used the lattice determination of $f_K/f_\pi$ to
determine $|V_{us}|$ employing the relation 
  	\begin{equation}
   		\frac{\Gamma(K\to\mu\bar\nu_\mu (\gamma))}
	      	{\Gamma(\pi\to\mu\bar\nu_\mu (\gamma))}=
		{\frac{|V_{us}|^2}{|V_{ud}|^2}}
		\left({ \frac{f_K}{f_\pi}}\right)^2
		\frac{m_K(1-\frac{m_\mu^2}{m_K^2})}
		{m_\pi(1-\frac{m_\mu^2}{m_\pi^2})}\times
			0.9930(35)\,.
  	   \end{equation}
From the charged kaon and pion life-times and the leptonic partial widths in
\cite{PDBook} one determines
	\begin{eqnarray}
	 \Gamma(K\to\mu\bar\nu_\mu(\gamma))  &=&2.528(2)\times 10^{-14}{\rm MeV}\,,
		\\	
	 \Gamma(\pi\to\mu\bar\nu_\mu(\gamma))&=&3.372(9)\times 10^{-14}{\rm MeV}\,,
		\nonumber
	\end{eqnarray}
which then yields
	\begin{equation}
	{\frac{ |V_{us}|^2}{ |V_{ud}|^2}}
		{ \frac{f^2_K}{f^2_\pi}}=
		0.07602(23)_{\rm exp}(27)_{\rm RC}\,,
	\end{equation}
with errors from experiment and radiative corrections, respectively.
Combining the very precise result 
${ |V_{ud}|}=0.97377(11)_{\rm exp}(15)_{\rm nucl}(19)_{\rm RC}$\footnote{
A preliminary update for this number was given by Marciano at the
Kaon 2007 International Conference: 
${ |V_{ud}|}=0.97372(10)_{\rm exp}(15)_{\rm nucl}(19)_{\rm RC}$
\cite{Marciano_Kaon}
} 
\cite{Marciano:2005ec}
from nuclear $\beta$-decay 
with the prediction for $f_K/f_\pi$ from lattice computations one determines
$|V_{us}|$.
\begin{table}\label{tab:fKfpi_collabs}
\begin{center}
\begin{tabular}{lc@{\hspace{1mm}}c@{\hspace{1mm}}l@{\hspace{1mm}}r@{\hspace{4mm}}c@{\hspace{2mm}}c@{\hspace{1mm}}l@{\hspace{1mm}}l@{\hspace{1mm}}lll}
\hline\hline\\[-5mm]
group	  &&$N_f$&action		&$a$/fm&$a$ from&$Lm_\pi^\dagger$& $m_\pi$/MeV&
	\multicolumn{1}{c}{$f_K/f_\pi$}\\
\hline\hline\\[-5mm]
{\hspace{-1.5mm}\begin{tabular}{l}
	 QCDSF+\\[-2mm]
	UKQCD
 \end{tabular} }&{\cite{Schierholz}}  &2& clover (NP)	&$\gtrsim 0.06$&$r_0$&4.2&		$\gtrsim 300$&1.21(3)\\[-0mm]
{ ETM}&{ \cite{Tarantino,ETM:2007}}  &2& max. tmQCD		&$0.09$&$f_\pi$     &3.2&		$\gtrsim 290$&1.227(9)(24)\\[-0mm]
{ MILC }&{\cite{Aubin:2004fs,Bernard:2006wx,Bernard}}    &2+1  & ${\rm KS_{\rm MILC}^{\rm AsqTad}}$&    $\gtrsim0.06$&$f_\pi$&$4$&
	$\gtrsim 240$&$1.197(3)(^{+6}_{-13})$\\[-0mm]
{\hspace{-1.5mm}\begin{tabular}{l}HPQCD+\\[-2mm]
	UKQCD
	\end{tabular} }&{\cite{Follana:2007uv}} &2+1& $\rm KS^{\rm HISQ}_{\rm MILC}$	&    $\gtrsim0.09$&$\Upsilon$&3.8&	$\gtrsim 250$&1.189(7)\\ [-0mm]
{ NPLQCD }&{\cite{Beane:2006kx}} &2+1&$\rm KS_{\rm MILC}/DWF$ 
	&   0.13&$r_0$&3.7&	$\gtrsim 290$&$1.218(2)(^{+11}_{-24})$\\[-0mm]
{\hspace{-1.5mm}\begin{tabular}{l} RBC+\\[-2mm]UKQCD\end{tabular} }&{\cite{Allton:2007hx,Scholz}} &2+1& DWF		&    0.11&$\Omega^-$&4.6	&	$\gtrsim 330$&1.205(18)\\[-0mm]
{ PACS-CS }&{\cite{Kuramashi}}  &2+1& clover (NP)	&    0.09&$\phi$&3	&	$\gtrsim 210$&{1.219(26)}\\[-0mm]
\hline\\[-6mm]
\multicolumn{9}{c}{\tiny$^\dagger$ lightest pion; NP:  non-perturbatively 
improved; max. tmQCD: maximally twisted mass QCD
}\\
\hline
\hline
\end{tabular}
\end{center}\mbox{}\\[-12mm]
\caption{Parameters of gauge configurations from which
	$f_K/f_\pi$ has been determined. Errors on the results 
	are either statistical and
	systematic or the combined error.}\label{tab:fKfpi_params}
\end{table}
Table \ref{tab:fKfpi_params} summarises the basic parameters and results 
for the 
gauge field ensembles from which $f_K/f_\pi$ has recently been determined. 
I begin with some comments on the treatment of the systematic error 
which is summarised in table \ref{tab:fKfpi_systematics}:
\begin{table}
\begin{center}
\footnotesize
\begin{tabular}{lcccll}
\hline\hline&&&&\\[-4mm]
group	  &$am_l$-extrapolation	&$a$	&FVE\\
\hline\hline\\[-4mm]
{ QCDSF+UKQCD}  &linear extrapolation&2 values of $a$&$\chi$PT\\[0mm]
\hline\\[-4mm]
{ ETM}  &\hspace*{-3mm}
	\begin{tabular}{c}
		NLO$\chi$PT+NNLO (analytic terms)\\[-1mm]
		and also polynomial
	\end{tabular}&-&$\chi$PT\\[0mm]
\hline\\[-4mm]
{ MILC}       &\multicolumn{2}{l}{rS$\chi$PT: first $\rm NLO+NNLO$ analytic
	\,terms on reduced set}&{2 volumes}\\[-1mm]
&\multicolumn{2}{c}{then include NNNLO-analyt. terms and}
		& rS$\chi$PT\\[-1mm]
&\multicolumn{2}{c}{include also larger values of $am_l$, 4 values of $a$}\\[-0mm]
\hline\\[-4mm]
{ HPQCD+UKQCD} & \multicolumn{2}{c}{NLO$\chi$PT+NNLO(analytic)}&$\chi$PT\\[-1mm]
		&\multicolumn{2}{c}{$a^2$ 
			(conventional+taste breaking)-terms} \\[-1mm]
&\multicolumn{2}{c}{3 values of $a$} \\[-0mm]
\hline\\[-4mm]
{ NPLQCD}    &NLO$\chi$PT+NNLO analytic terms &-&-\\[0mm]
\hline\\[-4mm]
{ RBC+UKQCD}  &NLO$\chi$PT$^{SU(2)\times SU(2)}$&-&2 volumes\\[0mm]
\hline\\[-4mm]
{ PACS-CS}  &NLO$\chi$PT&-&2 volumes, \cite{Colangelo:2005gd}\\[0mm]
\hline\\[-4mm]
\multicolumn{4}{c}{\tiny $\chi$PT: continuum chiral perturbation theory - 
	rS$\chi$PT: rooted staggered chiral perturbation theory 
		(including cut-off dependence)}\\
\hline\hline\\
\end{tabular}
\end{center}\mbox{}\\[-15mm]
\caption{Summary of how the systematic effects due to 
	the extrapolation in the quark mass, cut-off effects and 
	finite volume effects have been treated in the determination 
 	of $f_K/f_\pi$.}
\label{tab:fKfpi_systematics}
\end{table}

\textit{Finite volume effects}
were studied by all collaborations except NPLQCD by comparing results 
from lattices with different physical volumes but otherwise
constant physical parameters (MILC, RBC+UKQCD, PACS-CS) 
and/or using NLO chiral perturbation theory 
(QCDSF+UKQCD, ETM, MILC, HPQCD+UKQCD)
\cite{Gasser:1987zq,Becirevic:2003wk}.
I should comment that the first method  is clearly 
preferred and that 
values of $m_\pi L\approx 3$ very likely lead to finite volume effects 
that are not negligible.
In addition to comparing results from two volumes for some simulation
points PACS-CS estimated finite volume effects using the approach by 
Colangelo-D\"urr-Haefeli \cite{Colangelo:2005gd}. It is based on
L\"uscher's idea to
express finite volume effects of pion (kaon)
masses in terms of the $\pi\pi(K)$ scattering 
phase shift \cite{Luscher:1985dn} and predicts larger finite volume effects
than NLO $\chi$PT. Since this approach depends 
only indirectly 
on chiral perturbation theory through the estimate of the scattering
phase shift the predictions are expected to be more accurate. 

\textit{chiral extrapolation}:
UKQCD+QCDSF don't see a curvature in their data for $f_K/f_\pi$ and therefore
linearly extrapolate to the physical point.
Staggered fermions come in tastes and at finite lattice spacing the
taste symmetry is broken and the corresponding pion spectrum is
non-degenerate. The continuum chiral effective Lagrangian therefore does not
correctly describe the spectrum of staggered lattice QCD.
MILC, who are using AsqTad improved staggered valence and
sea quarks, therefore employ $SU(3)\times SU(3)$
rooted staggered partially quenched chiral perturbation theory 
\cite{Aubin:2003uc,Aubin:2003mg} which simultaneously 
describes the approach to the chiral and the continuum limit.
They first fit the NLO expression including NNLO analytic terms to 
a reduced data set including only lighter data points and
subsequently extend
the fit to the heavier data points by including NNNLO analytic terms
\cite{Bernard}.
HPQCD+UKQCD are simulating the partially quenched theory 
using HISQ \cite{Follana:2006rc} valence quarks on the MILC staggered sea
\cite{MILCconfigs,Bernard}.
Arguing that for this setup the taste splitting is
reduced sufficiently, they rely on NLO $SU(3)\times SU(3)$
continuum chiral perturbation theory
with added NNLO analytical and cut-off terms. In their recent publication
\cite{Follana:2007uv} HPQCD+UKQCD  
quote the smallest error for $f_K/f_\pi$ among all the results presented here
and it will be interesting to compare the details of their chiral and
continuum extrapolation once they are accessible in an upcoming
publication.
ETM, MILC, HPQCD+UKQCD and NPLQCD also use $SU(3)\times SU(3)$ 
NLO continuum chiral perturbation
theory adding NNLO analytical terms to the fit-ansatz in order to improve
the fit quality; PACS-CS is using NLO only.
RBC+UKQCD use a different strategy. Assuming that the strange quark mass
is too heavy to be properly described by $SU(3)$ chiral perturbation theory
and $m_s\gg m_{u,d}$ they use $SU(2)\times SU(2)$ chiral perturbation theory
fits for the kaon sector \cite{Sharpe:1995qp}, to describe the light quark mass
dependence of the decay constants at fixed strange quark mass. 
The results obtained in this way are then interpolated to the
physical point of the strange quark mass.

\textit{cut-off-effects}: Apart from MILC and HPQCD+UKQCD who are doing a 
combined chiral and continuum extrapolation
only QCDSF+UKQCD assess cut-off effects. The latter do not see any
scaling violations in their data. Some of the remaining collaborations will
supplement their current results with results on a finer lattice and/or rely
on the crude estimates for cut-off effects to be of order 
$(a\Lambda_{\rm QCD})^2$ in the $O(a)$-improved theory.
\subsection{$|V_{us}|$ from semi-leptonic decays}\mbox{}\\[-8mm]\indent\label{sctn:Kl3semilep}
The Standard Model expectation for the $K\to \pi l \nu$ ($K_{l3}$) 
semi-leptonic decay rate is \cite{Leutwyler:1984je}
\begin{equation}
   \Gamma_{K\to\pi l\nu}=C_K^2\frac{G_F^2 m_K^5}{192\pi^3} 
		\,{ I}\, { S_{\rm EW}}[1+
		{ 2\Delta_{SU(2)}} + 
		2{\Delta_{\rm EM}}]
		{|V_{us}|}^2 |{ f_+^{K\pi}(0)}|^2\,.
\end{equation}
Here, $C_K^2=1/2(1)$ is the Clebsch-Gordan coefficient for the neutral 
(charged) kaon
decay and $I$ is the phase space integral which is typically determined from
the shape of the experimentally measured form factor \cite{Leutwyler:1984je}. 
$S_{\rm EW}$ is the
electro-weak short distance correction and $\Delta_{SU(2)}$ and 
$\Delta_{\rm EM}$ are $SU(2)$-isospin breaking and electromagnetic corrections,
respectively.
The non-perturbative contribution to the process is given in terms of the
form factor $f_+^{K\pi}(0)$ defined through the QCD matrix element of
the vector current $V_\mu=\bar s \gamma_\mu u$ between the kaon and the pion,
\begin{equation}
	\langle \pi(p_\pi)|V_\mu(0)|K(p_K)\rangle =
        { f_+^{K\pi}(q^2)}(p_K+p_\pi)_\mu
        + f_-^{K\pi}(q^2)(p_K-p_\pi)_\mu\,,
\end{equation}
where $q_\mu=(p_K-p_\pi)_\mu$ is the momentum transfer.
Recently the FlaviaNet Kaon Decay Working Group determined the very accurate
value 
$|{ V_{us}}{ f_+^{K\pi}(0)}|=0.21673(46)$ \cite{Moulson:2007fs}
which can only be fully appreciated  for a determination 
of $|V_{us}|$ if lattice computations
 of $f_+^{K\pi}(0)$ reach sub-percent precision.

In chiral perturbation theory $f_+^{K\pi}(0)$ is expanded as 
\begin{equation}
f_+^{K\pi}(0)=1+f_2+f_4+\dots\,,
\end{equation}
where $f_2$, which corresponds to the leading chiral correction,
is fully determined in terms of the meson masses \cite{Gasser:1984ux} 
and takes the value $f_2^{\rm phys}=-0.023$ at the physical point. 
Relevant for lattice
computations are also the more recent evaluations of $f_2$ in 
partially quenched chiral perturbation theory and the evaluation 
of finite volume corrections \cite{Becirevic:2004ya,Becirevic:2005py}.
A first estimate of the higher order corrections was given in 
\cite{Leutwyler:1984je}
and more recently estimates for $f_4$ were given in 
\cite{Post:2001si,Bijnens:2003uy,Cirigliano:2005xn}. 
Thus, relying on our knowledge of $f_2$ in chiral perturbation theory, 
on the lattice one computes only the corrections,
\begin{equation}
	\Delta f(m_K,m_\pi)  = f_+(0,m_K,m_\pi)-(1+f_2(m_K,m_\pi))\,.
\end{equation}
Considering for a moment the first estimate $\Delta f=-0.016(8)$ 
for physical pion and kaon masses by Leutwyler
and Roos in 1984
\cite{Leutwyler:1984je}, it is clear that 
a precision of 30-40\% for $ \Delta f$ is sufficient to reach 
sub-percent accuracy for $f_+^{K\pi}(0)$.

During the last three years
a number of collaborations have computed 
this quantity using dynamical fermions. However, in terms of controlling
systematic uncertainties the project by
the RBC+UKQCD collaboration which use 2+1 flavours of dynamical 
domain wall fermions
with pion masses down to 330 MeV is the current  
 state of the art \cite{Antonio:2007mh,Juettner_Kaon,James} (cf. the summary
of recent simulations in table \ref{tab:Kl3_prev}).
At this conference also ETM \cite{Tarantino,Simula}
and QCDSF \cite{Morozov} indicated that they will compute the
$K_{l3}$ form factor with twisted mass fermions and clover
improved Wilson fermions, respectively. QCDSF gave a first estimate in their
proceeding \cite{Morozov} and one has to wait for their final error analysis 
and the extension of their analysis to lighter pion masses and one 
coarser and one finer lattice spacing.
\begin{table}
\begin{center}
\vspace{3mm}
{\small
\begin{tabular}{l@{\hspace{1mm}}c@{\hspace{1mm}}c@{\hspace{1mm}}c@{\hspace{1mm}}c@{\hspace{1mm}}c@{\hspace{1mm}}c@{\hspace{1mm}}l}
\hline\hline\\[-4mm]
group     &&$N_f$&action         &$a$/fm&$L$/fm&         $m_\pi$/MeV&
		$f_+^{K\pi}(0)$\\
\hline\hline\\[-4mm]
{JLQCD }&{\cite{Tsutsui:2005cj}}      &2  & clover (NP)         &     0.09&1.8&         $\gtrsim 550$&0.967(6)\\[-0mm]
{RBC }&{\cite{Dawson:2006qc}}      &2  & DWF           &    0.12&2.5&          $\gtrsim 490$&0.968(9)(6)\\[-0mm]
{QCDSF }&{\cite{Morozov}}      &2  & impr. Wilson           &    0.08&1.9&          $\;\;\;\; 592$&$0.9647^\ast$\\[-0mm]
{\hspace{-2mm}\begin{tabular}{l}Fermilab,\\[-1mm]HPQCD, MILC\end{tabular} }&{\cite{Okamoto:2004df}$^\dagger$}&2+1& \hspace{-2mm}\begin{tabular}{l}Wilson $d$-quark\\[-1.5mm]
  impr. stag. $u$ \& $s$\end{tabular}
        &  $\ddagger$  &$\ddagger$ &$\ddagger$      &0.962(6)(9)    \\[-0mm]
{RBC+UKQCD }&{\cite{Antonio:2007mh,Juettner_Kaon,James}}  &2+1& DWF         &    0.11&1.8,2.8&              $\gtrsim 330$&0.964(5)\\[-0mm]
\hline\\[-5mm]
\multicolumn{8}{c}{\tiny $^\dagger$ computed $f_0^{K\pi}(q^2_{\rm max})$ and then
used slope of form factor from experiment for extrapolation to $q^2=0$ 
	("exploratory study");}\\[-2mm]
\multicolumn{8}{c}{\tiny $^\ast$ Error analysis not yet finished }\\[-2mm]
\multicolumn{8}{c}{\tiny $^\ddagger$
The proceeding in which the result was published does not contain
this information}\\
\hline\hline
\end{tabular}}
\end{center}\mbox{}\\[-12mm]
\caption{Lattice computations of the $K_{l3}$ form factor 
using dynamical fermions.  Errors on the results 
	are either statistical and
	systematic or the combined error.}\label{tab:Kl3_prev}
\end{table}

The technique for the high precision calculation of $f_+^{K\pi}(0)$ has 
been set out in \cite{Becirevic:2004ya} and consists of a 3-step procedure:\\[-8mm]
\bi
 \item[1)] one first computes $f_0^{K\pi}(q_{\rm max}^2) = f_+^{K\pi}(q_{\rm max}^2)+q^2_{\rm max}/(m_K^2-m_\pi^2)f_-(q^2_{\rm max})$, where 
	$q_{\rm max}^2=(m_K-m_\pi)^2$,  from
        \begin{equation}\label{eq:kl3_rat1}
         \frac{
         \langle \pi|V_0|K\rangle \langle K|V_0|\pi \rangle}
         {\langle \pi|V_0|\pi\rangle \langle K|V_0|K\rangle }=
         \frac{(m_K+m_\pi)^2}{4m_K m_\pi}
                \left( f_0(q^2_{\rm max},m_K,m_\pi)\right)^2\,.
        \end{equation}
	The l.h.s. is obtained from ratios of suitable Euclidean
	three-point functions at large values of the Euclidean time.
	\begin{figure}
	 \begin{center}
	  \begin{minipage}{.48\linewidth}
	\small
	\begin{tabular}{l@{\hspace{0mm}}c@{\hspace{0mm}}|c@{\hspace{1mm}}c@{\hspace{1mm}}c@{\hspace{1mm}}c@{\hspace{1mm}}|c@{\hspace{1mm}}c@{\hspace{1mm}}c}
	\hline\hline&&&&&&&&\\[-5mm]
	        &&\multicolumn{4}{c|}{$q^2$-fits}        
		&\multicolumn{3}{c}{chiral fits}\\[-1mm]
	        &&lin&quad&pole&z&const&lin&quad\\
	\hline\hline&&&&&\\[-4mm]
	JLQCD &\cite{Tsutsui:2005cj}&   &\bdot& & &              &\bdot\\[-0mm]
	RBC   &\cite{Dawson:2006qc}&$\circ$&$\circ$&\bdot &&$\circ$&\bdot&$\circ$\\[-0mm]
	\hspace{-2mm}\begin{tabular}{l}Fermilab\\[-1.5mm]FNAL,MILC\end{tabular}
		 &\cite{Okamoto:2004df}&& &$\,\,\,\bullet^\dagger$&&&\bdot\\[-0mm]
	RBC/UKQCD &\cite{Antonio:2007mh}&$\circ$&$\circ$&\bdot &$\circ$&$\circ$&\bdot&$\circ$ \\
	\hline
\multicolumn{9}{c}{\tiny
		\bdot\, = adopted; $\circ$ = tested; }\\[-3mm]
\multicolumn{9}{c}{\tiny
$\dagger$ = 
	\hspace{-2.5mm}
	 \begin{tabular}{l}
		pole mass from experiment $\to$ extrapolate
	 	to $f_0(0)$ from $f_0(q^2_{\rm max})$\\
	\end{tabular}}\\
	\hline\hline
	\end{tabular}
	  \end{minipage}
	  \begin{minipage}{.48\linewidth}
		\hspace{8mm}
	   \epsfig{scale=.29,file=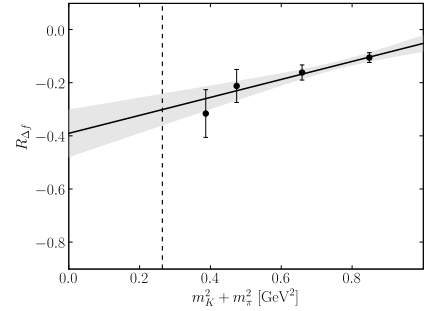}
	  \end{minipage}
	 \end{center}\mbox{}\\[-12mm]
	\caption{
	Left: Summary of how systematic in chiral extrapolation and
		$q^2$-interpolatoin was assessed; right: chiral extrapolation of
	$R_{\Delta f}$  (r.h.s. plot taken from
	\cite{James}).
	}\label{fig:kl3_rat1}
	\end{figure}
	Since both mesons are at rest and 
	due to cancellations of correlations and also the cancellation of
	the vector current renormalisation constants in the ratio  
	in (\ref{eq:kl3_rat1}), $f_0^{K\pi}(q_{\rm max}^2)$ 
	can be determined with very high precision.\\[-8mm]
 \item[2)] One then determines $f_0^{K\pi}(q^2)$ at various other
        values of $q^2<q^2_{\rm max}$ from similar ratios of 3-point functions
	by inducing 
	Fourier momenta ($|\vec p_{K,\pi}|=$$2\pi/L,\sqrt{2}\,2\pi/L$)
	into the initial and/or
 	final state. In order to interpolate the form factor
	to $q^2=0$ the collaborations have used first or second order
	polynomial ans\"atze in $q^2$, a pole dominance ansatz
	$f_{0,\rm pole}(q^2) = \frac{f_0(0)}{1-q^2/M^2}$ or the 
	$z$-fit 
	(polynomial\, with\, improved\, convergence\, \cite{Hill:2006bq})
	as summarised in the table in figure~\ref{fig:kl3_rat1}.\\[-8mm]
 \item[3)] In the last step the data for $f_+^{K\pi}(0)=f_0^{K\pi}(0)$ has to 
	be extrapolated to the physical value of the quark masses and 
	it is common practice to use the ratio
       \begin{equation}\label{eq:RDeltaf}
         R_{\Delta f}\equiv\frac{\Delta f }{(m_K^2-m_\pi^2)^2}=
                \frac{f_0^{K\pi}(0)-(1+f_2(m_K,m_\pi))}{(m_K^2-m_\pi^2)^2}\,,
        \end{equation}
        with a polynomial ansatz constant, 
	linear or quadratic in ($m_K^2+m_\pi^2$).
	Again the table in figure  \ref{fig:kl3_rat1} 
	summarises which ansatz has been
 	studied by the various collaborations.
	An example for the extrapolation with a linear fit is shown in 
     	in the plot in figure \ref{fig:kl3_rat1}
	which is taken from James Zanotti's talk \cite{James}.
\ei
Both the $q^2$-interpolation and the chiral extrapolation are 
based on phenomenological fit-ans\"atze and are thus sources
of systematic uncertainties which can be estimated in terms of  the spread
of results between the different ans\"atze.
The RBC+UKQCD-collaboration \cite{James} has observed a reduction in the final 
error when combining the
$q^2$-interpolation and the chiral extrapolation into one global
fit. In the next section
we will briefly introduce a new technique \cite{Boyle:2007wg}
that may entirely remove the uncertainty due
to the $q^2$-interpolation.
Before let me make some comments:\\[-8mm]
\bi
 \item It would be interesting  to have the expression in
	chiral perturbation theory of the 
	$K_{l3}$ form factor  to order $p^6$ in \cite{Bijnens:2003uy} 
	in terms of the quark masses.
	This would on the one hand allow
	to understand the slight tension between $\chi$PT and the lattice 
	results which I will comment on later.
	 On the other hand
	lattice results could then also be used to constrain or even determine
	the low energy constants appearing in these expressions.\\[-8mm]
 \item Taking the preliminary result by RBC+UKQCD \cite{Antonio:2007mh} as the
	current bench-mark, the contributions to the overall error
	on $\Delta f$ from the lattice decomposes in the 
	following way\footnote{\,In order to disentangle the various 
	contributions to the systematic
	error I here quote the results from the individual 
	$q^2$-interpolation and chiral extrapolation in \cite{James}.}:
        \begin{equation}
        \begin{array}{rcccc}
        \Delta f=-0.0161&(45)^{\rm stat.}&(15)^{\rm \chi}&(16)^{q^2}&(8)^{a}\\[0mm]     
                        &30\%&9\%&10\%&4\%\,,
        \end{array}
        \end{equation}
	for the statistical error, error due to the chiral extrapolation,
	error due to the $q^2$-interpolation and error due to 
	the finite lattice-cut-off, respectively.
	In the following we will briefly discuss two recent developments which
	should allow to further reduce the error due to the $q^2$-interpolation
	on the one hand and the statistical error on the other hand.
\ei
\subsection{$K_{l3}$ with partially twisted boundary conditions}\mbox{}\\[-8mm]\indent
Using twisted boundary conditions for the valence quark fields,
i.e.  $q(x+L\hat{k})=e^{i\theta_k}q(x)$ for $k=1,2,3$ and leaving 
the sea quark's boundary conditions untouched (\textit{partially twisted 
boundary conditions}) \cite{Bedaque:2004kc,Sachrajda:2004mi,Bedaque:2004ax},
 one can tune the momenta of hadrons in dynamical simulations
in a finite box continously.
For example the dispersion relation for a charged pion 
then follows 
\begin{equation}
E_{\pi^\pm}=\sqrt{m_{\pi^\pm}^2+ (\vec p_{\rm lat}
        {-{{1\over L}(\vec\theta_u-\vec\theta_{ d})}})^2}\,,
\end{equation}
where $\vec\theta_u$ and $\vec\theta_d$ are the twisting angles 
of the valence  up and down quarks, respectively, and $\vec p_{\rm lat}$ is
the conventional Fourier momentum.
This relation was confirmed numerically 
in \cite{Flynn:2005in}.
Twisted boundary conditions were also tested in the quenched theory
\cite{deDivitiis:2004kq} where they have subsequently been been 
applied to
study the $q^2$-dependence of the $K_{l3}$ form factor using techniques 
similar to the 3-step procedure detailed in the previous section 
\cite{Guadagnoli:2005be}.
An approach which goes beyond that study was developed in 
\cite{Boyle:2007wg} where the matrix element 
\begin{equation}\label{eq:tw_ff}
\langle \pi(p_\pi)|V_4(0)|K(p_K)\rangle =
              {f_+^{K\pi}(0)}(E_K+E_\pi)
           + f_-^{K\pi}(0)(E_K-E_\pi)\,,
\end{equation}
of the 4th component of the vector current
is evaluated for two choices of the kinematics, namely
        \begin{equation}\label{eq:twists}
         \begin{array}{llcccc}
             1)&\langle \pi(0)|V_4|K(\vec \theta_K)\rangle|_{q^2=0}:& |\vec{\theta}_K| =
                  L\sqrt{({m_K^2+m_\pi^2 \over 2m_\pi})^2 - m_K^2}
             &\textrm{and}&\vec{\theta}_\pi=\vec{0}\,,\\[2mm]
             2)&\langle \pi(\vec \theta_\pi)|V_4|K(0)\rangle|_{q^2=0}:& |
                \vec{\theta}_\pi| =L
                 \sqrt{({m_K^2+m_\pi^2 \over 2m_K})^2 -
                 m_\pi^2}&\textrm{and}
             &\vec{\theta}_K =\vec{0}\,.\\[2mm]
         \end{array}\nonumber
        \end{equation}
The form factor $f_+^{K\pi}(0)$ in eq.~(\ref{eq:tw_ff})
can directly be extracted from a linear 
combination of the matrix element in the kinematical situations
1) and 2) in (\ref{eq:twists}), 
thus entirely avoiding the $q^2$-interpolation and difficult fits to 
ratios of correlation functions involving the spatial components of the 
vector current, which are necessary in the 3-step procedure lined out
above. 
The statistical errors
of the results from the new approach are comparable to the errors
using the conventional approach. 
\subsection{$K_{l3}$-decays using propagators from stochastic sources}
\begin{figure}
\begin{center}
        \epsfig{scale=.22, file=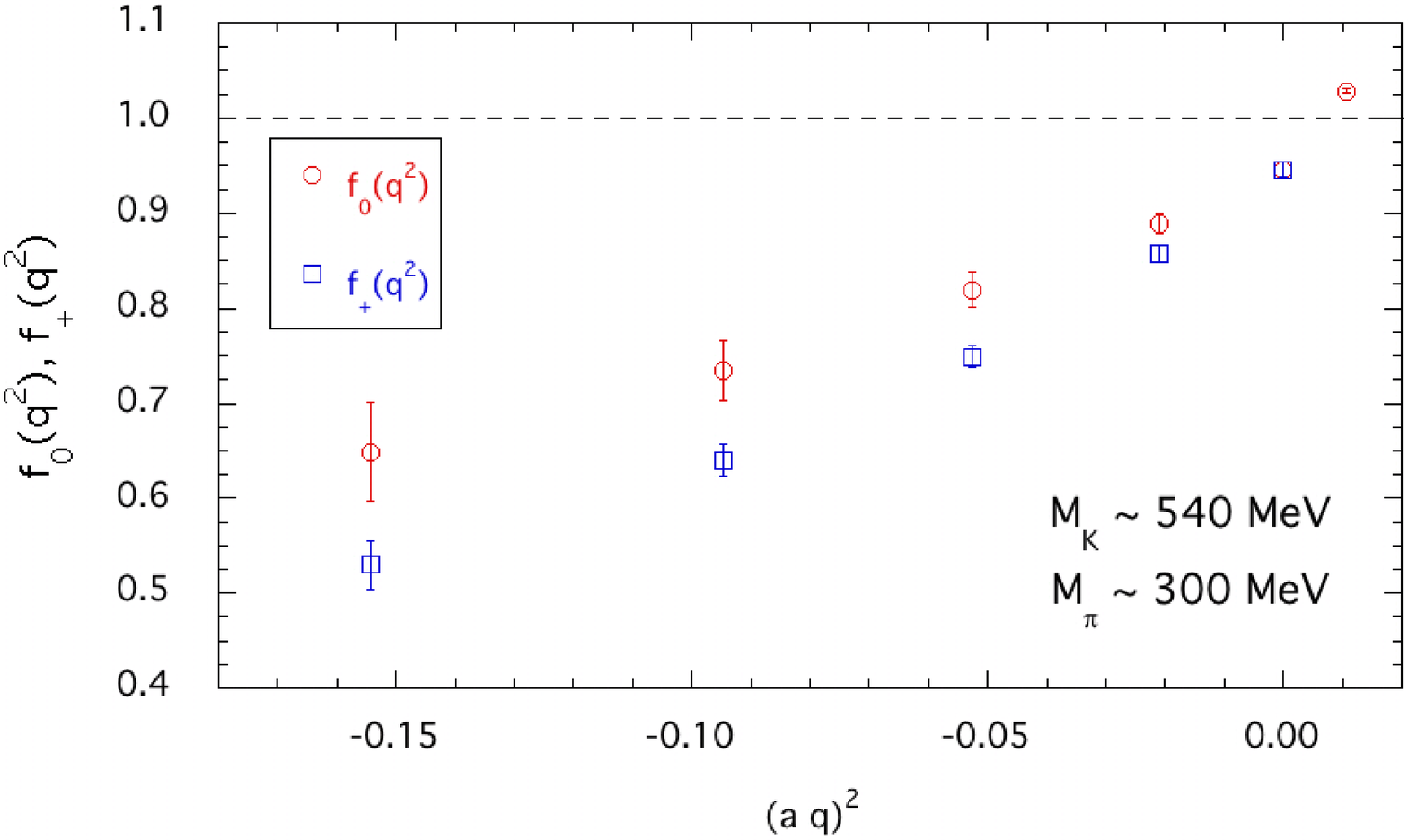}
	\hspace{8mm}
        \epsfig{scale=.22, file=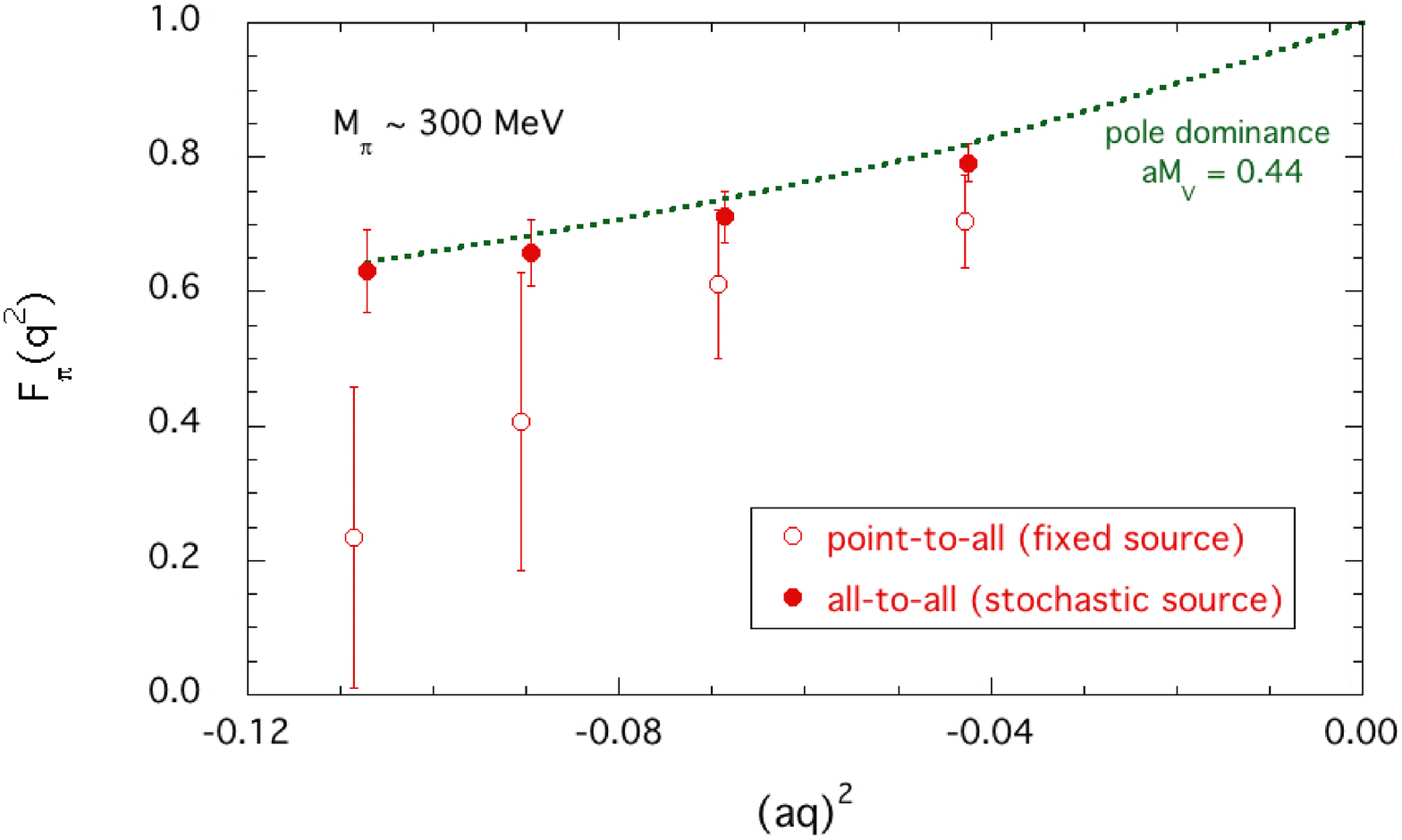}
	\mbox{}\\[-2mm]
 	\caption{Left: $f_{0}^{K\pi}(q^2)$ with all-to-all propagators; right:
		the pion form factor $f_\pi(q^2)$ constructed from 
		 point-to-all and all-to-all propagators, respectively
		\cite{Simula}.
		}
	\label{fig:Kl3_all-to-all}
\end{center}
\end{figure}

\mbox{}\\[-8mm]\indent
Instead of constructing the three
point functions relevant to the determination of the $K_{l3}$-form factor
from  point-to-all propagators,
the ETM collaboration \cite{Simula} uses all-to-all propagators 
generated with the one-end-trick
\cite{Foster:1998vw,McNeile:2006bz}, thus
 gaining a volume averaging of the propagator source.
The pion and kaon momenta in these simulations 
are induced using partially twisted boundary
conditions \cite{Bedaque:2004kc,Sachrajda:2004mi,Bedaque:2004ax} in 
combination with the 3-step procedure lined 
out in section \ref{sctn:Kl3semilep}. 
Preliminary results for this approach were shown in
Simula's talk at this conference (cf. l.h.s. of 
 figure~\ref{fig:Kl3_all-to-all}). 
Since no data for comparison with results
from point-source propagators were available for 
$f_+^{K\pi}(q^2)$ and $f_0^{K\pi}(q^2)$, one gains some feeling for the 
improvement by looking at Simula's 
plot of the pion vector form factor $f^{\pi\pi}(q^2)$ on the r.h.s. of 
figure~\ref{fig:Kl3_all-to-all}. 
\subsection{Summary for $|V_{us}|$}
\begin{figure}
 \begin{center}
  \begin{minipage}{.4\linewidth}
\hspace{1.5mm} \epsfig{scale=.22,angle=-90,file=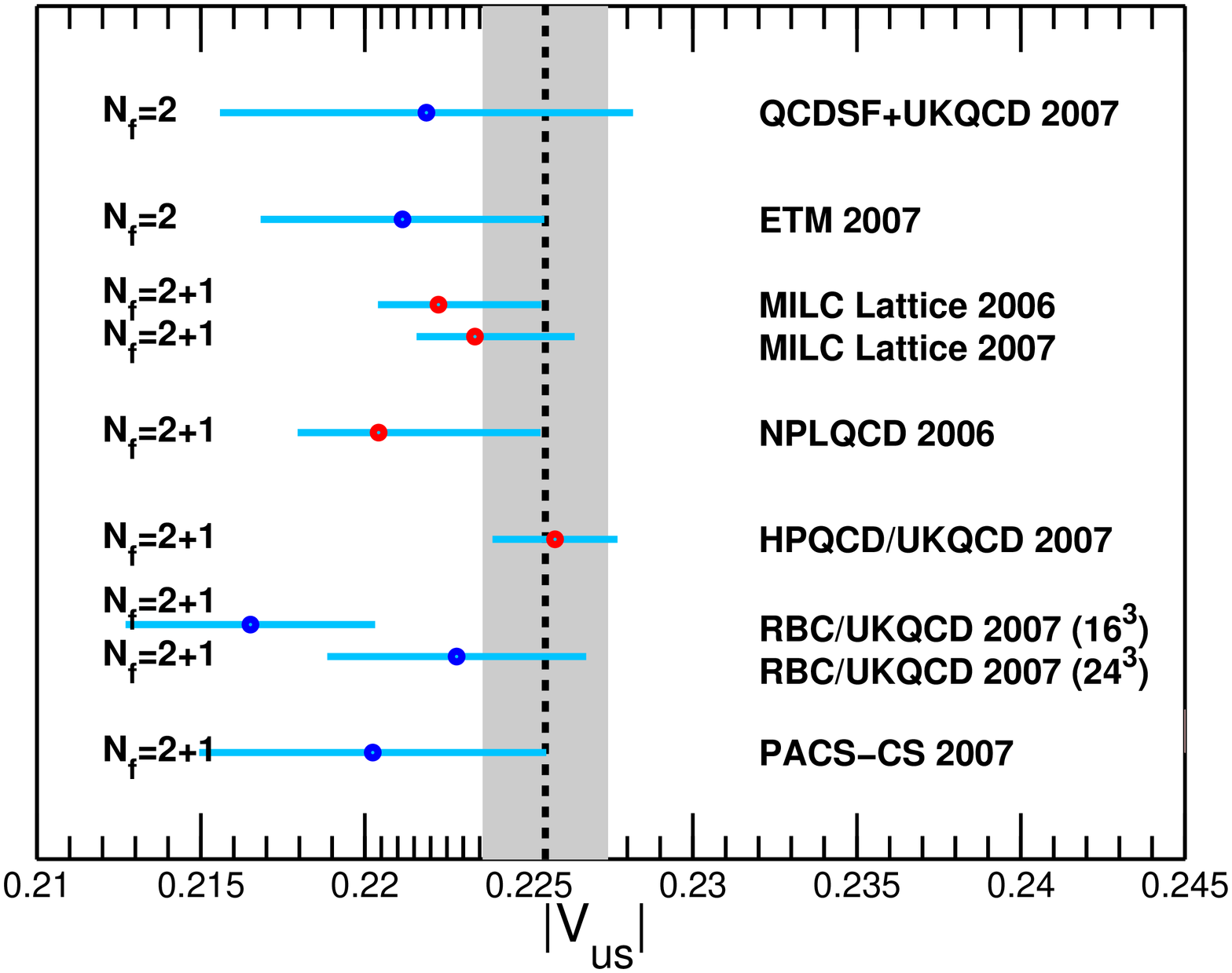}\\
        \begin{rotate}{90}{\hspace{5mm}from $K_{l2}$}\end{rotate}
\hspace{1mm} \epsfig{scale=.22,angle=-90,file=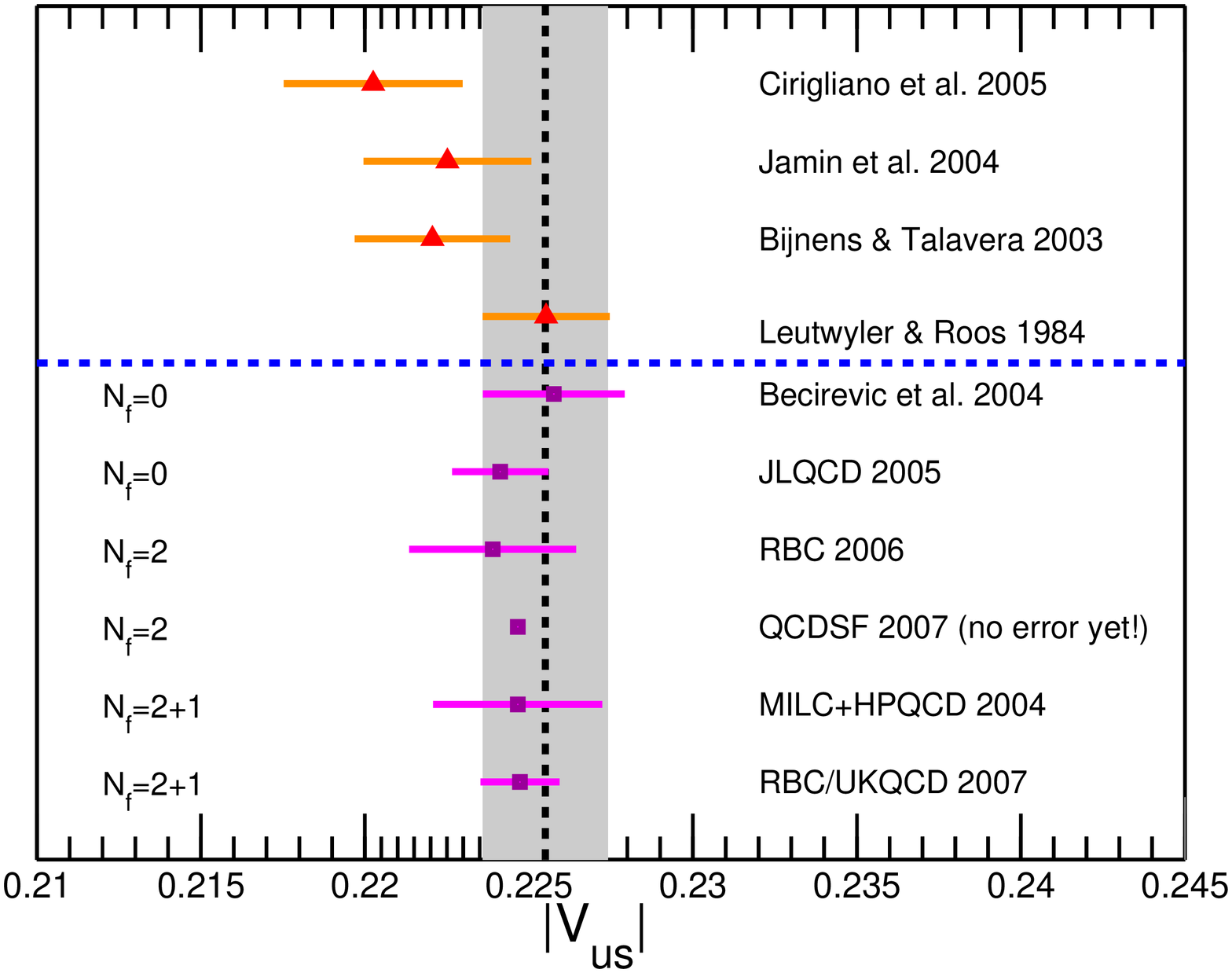}\\
	\begin{rotate}{90}{\hspace{7mm}from $K_{l3}$}\end{rotate}
  \end{minipage}
  \begin{minipage}{.48\linewidth}
         \epsfig{scale=.35,file=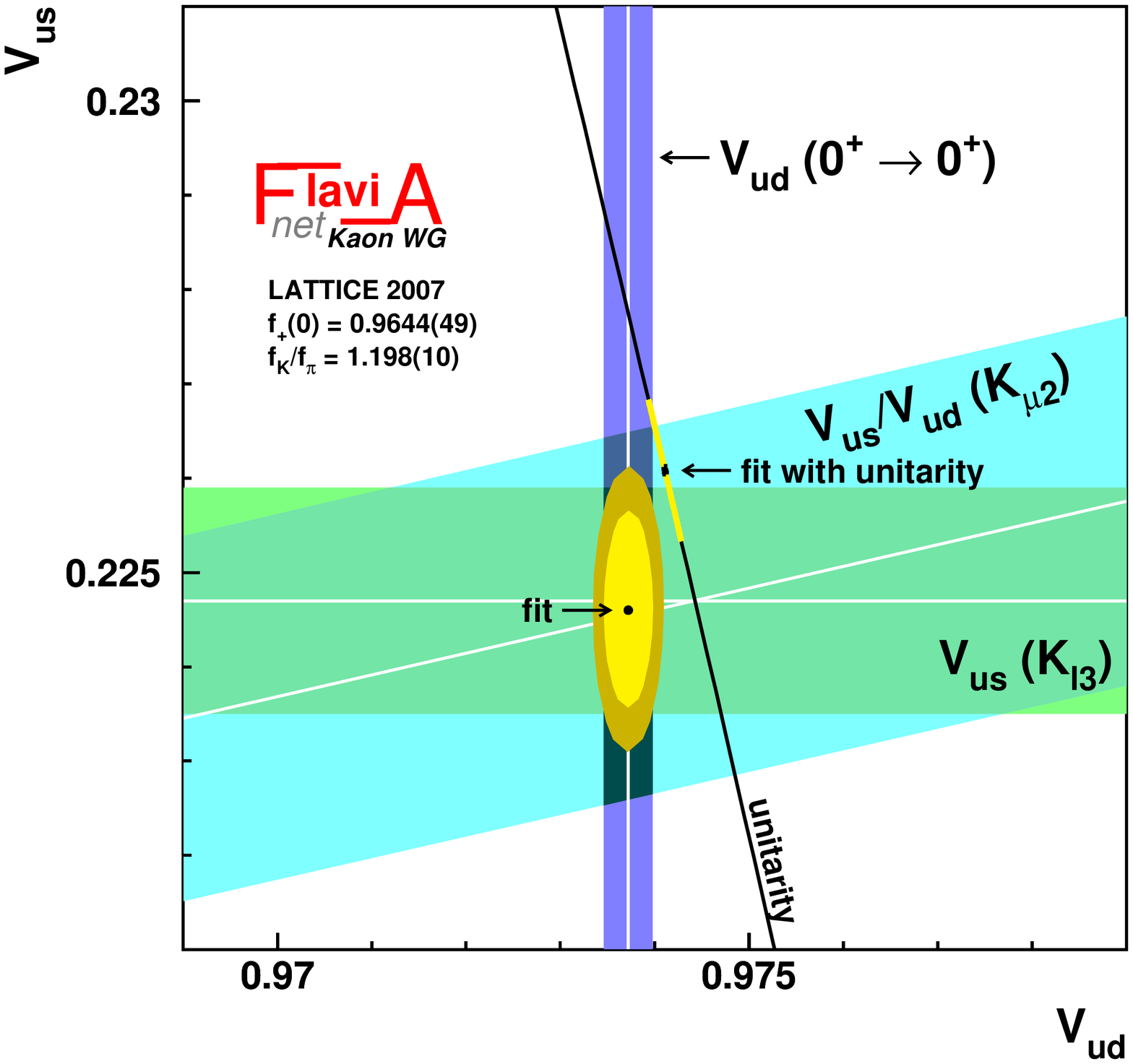}\\[-6mm]

  \end{minipage}
 \end{center}\mbox{}\\[-15mm]
\caption{Left: Comparison of results for the CKM-matrix element $|V_{us}|$ from
	lattice computations of the 
	leptonic and semi-leptonic kaon decays. In the lower plot also
	results from chiral perturbation theory are shown. Right: 
	Compilation of the estimates (\protect\ref{eqn:bestKl3}) and 
	(\protect\ref{eqn:bestKl2})
	into a
	combined fit by the FlaviaNet Kaon Decay WG
	(like in \cite{Moulson:2007fs})
	 to assess whether the first row of the CKM matrix
	fulfils the unitarity constraint.}
\label{fig:Vus_summary}
\end{figure}

\mbox{}\\[-8mm]\indent
Figure \ref{fig:Vus_summary} shows a comparison of recent results for $|V_{us}|$
from both leptonic and semi-leptonic kaon decays.
The grey band in the plot gives the result for $|V_{us}|$ which one gets
using Leutwyler and Roos' result $f_+^{K\pi}(0)=0.961(8)$ 
\cite{Leutwyler:1984je} and
$|{ V_{us}}{ f_+^{K\pi}(0)}|=0.21673(46)$ by
the FlaviaNet Kaon Decay Working Group \cite{Moulson:2007fs}. The lattice results
for $K_{l3}$ currently support Leutwyler and Roos' prediction.
Both chiral perturbation theory for $f_+^{K\pi}(0)$
\cite{Bijnens:2003uy,Jamin:2004re,Cirigliano:2005xn}  and
the lattice predictions for $f_K/f_\pi$ tend towards smaller
values for $|V_{us}|$, thus creating a slight tension.
We also see that 
results for the CKM-matrix element from lattice calculations of leptonic 
kaon decays are becoming competitive with results from calculations of 
semi-leptonic decays.

RBC+UKQCD currently have the best control of systematic 
effects in the calculation of $f_+^{K\pi}(0)$ \cite{Peter,Antonio:2007mh,James,Kl3-upcoming} and I take
their result as the current best estimate,
\begin{equation}\label{eqn:bestKl3}
\begin{array}{lcrcl}
f_+^{K\pi}(0)&=&0.964(5)\,\longrightarrow|V_{us}|=0.2247(12)\,.
\end{array}
\end{equation} 
Due to the preliminary status of some of the central values and error bars
 for $f_K/f_\pi$ in table 
\ref{tab:fKfpi_params} it is difficult to compute a world average with
a reliably estimated error. As the central value for the current best 
estimate I suggest the weighted average 
over all results in table \ref{tab:fKfpi_params}
assuming Gaussian and un-correlated errors\footnote{In 
	the case of assymmetric error
	bars I have shifted the central value of the result and corrected the 
	error correspondingly to be symmetric.}. 
MILC \cite{Bernard}
has carried out the most extensive simulation 
with a very detailed study of the chiral extrapolation, 
the finite-volume and the  cut-off  effects all within the frame work 
of rooted staggered chiral perturbation theory and I therefore attach
to the averaged central value their combined statistical and systematic
error,
\begin{equation}\label{eqn:bestKl2}
\begin{array}{rcl}
 f_K/f_\pi=1.198(10)\,\longrightarrow |V_{us}|=0.2241(24)\,,  
\end{array}
\end{equation}
which reveals  a slight tension with the experimental
value 1.223(12) \cite{PDBook}. Note that this average
is dominated by the MILC and HPQCD+UKQCD results with staggered fermions
which have very small errors. The average without the results based on  
staggered fermions would instead be 1.211(10).

The FlaviaNet Kaon Decay Working Group was so kind to provide a version
of their unitarity fit \cite{Moulson:2007fs,Mescia:2007ku}
assuming the above best estimates which is shown
in the r.h.s. plot in figure \ref{fig:Vus_summary}\,\footnote{
Many thanks to M. Antonelli, 
F. Mescia and  M. Moulson for generating this plot \cite{FNCKM}.}.
The blue vertical line represents Marciano's update
for $|V_{ud}|$ from nuclear $\beta$-decays \cite{Marciano_Kaon}. 
The horizontal band is the 
result for $|V_{us}|$ from RBC+UKQCD's lattice calculation of the
semi-leptonic form factor \cite{Peter,Antonio:2007mh,Kl3-upcoming} 
and the slightly tilted
horizontal band represents (\ref{eqn:bestKl2}).
The solid black line represents CKM first-row unitarity (neglecting
$|V_{ub}|$). This analysis which is partly based on preliminary results
 indicates a tension between the lattice results and CKM-unitarity.
With $|V_{ud}| = 0.97372(26)$ and 
$|V_{us}|=0.2246(11)$ one gets 
$|V_{ud}|^2+|V_{us}|^2-1=0.0014(7)$.
It will be extremely exciting to follow the development of this situation
as the various collaborations finalize their analysis and in particular 
theoretical errors are further reduced.
\section{$CP$-violation in kaon systems}\mbox{}\\[-8mm]\indent
The physical state
$K_L$ consists predominantly of the $CP$-odd $K_2$ and an admixture
of the $CP$-even $K_1$. The decay of the $K_2$ into a
$CP$-even two-pion state is called direct $CP$-violation and 
has been established experimentally in 1999 by NA48 and KTeV
\cite{Fanti:1999nm,AlaviHarati:1999xp}. Direct $CP$-violation which occurs
when the $K_1$ decays into the $CP$-even two-pion state has been 
established in 1964 by Cronin and Fitch \cite{Christenson:1964fg}.
Direct and indirect $CP$-violation have been studied on the lattice 
for many years and in the following I will discuss recent advances and 
results. 
\subsection{Indirect $CP$-violation - neutral kaon mixing}
\begin{figure}
 \begin{center}
 \epsfig{scale=.4,file=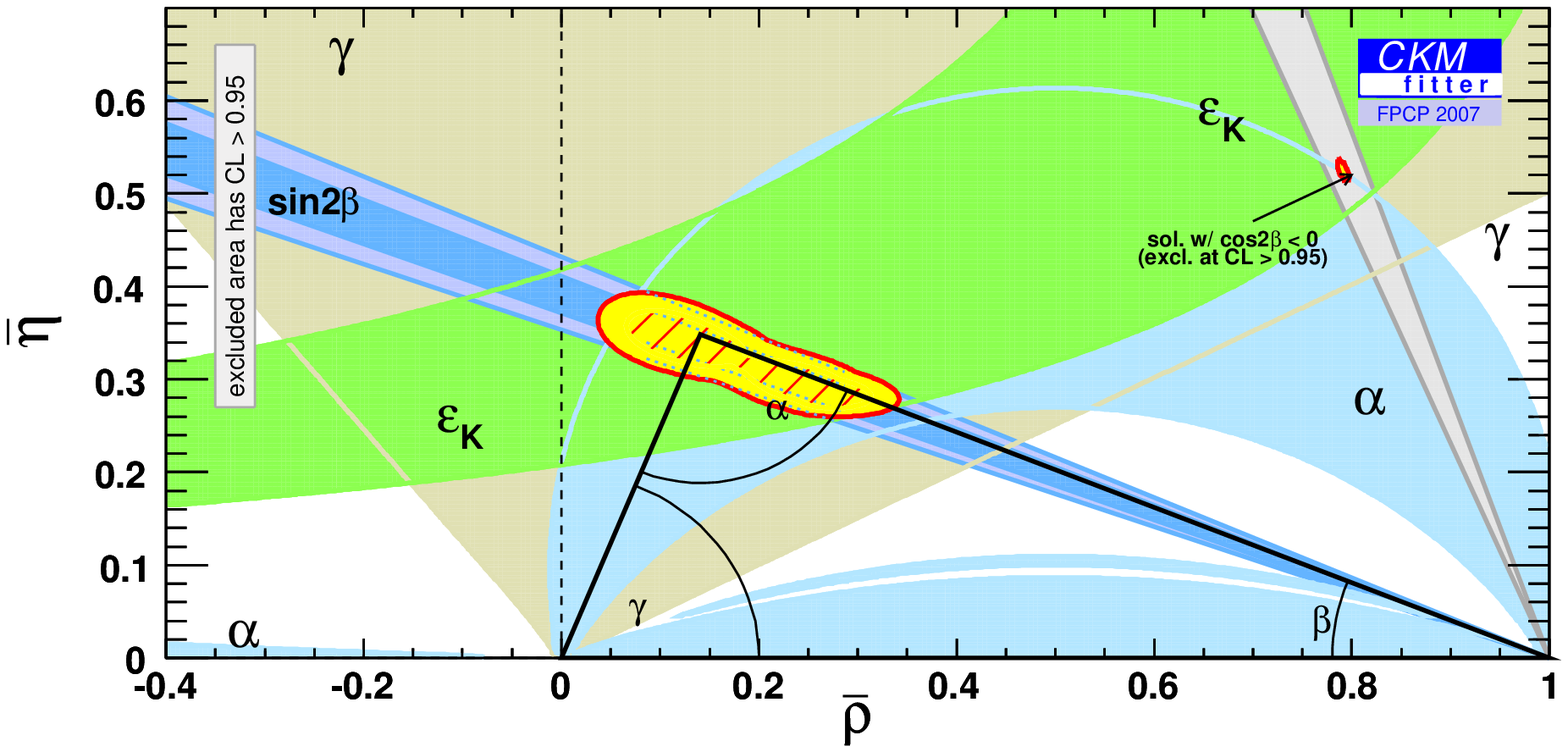}
 \end{center}
	\mbox{}\\[-18mm]
 \caption{Status of the CKM-triangle by the CKMFitter group 
\cite{Charles:2004jd}}\label{fig:CKM-triangle}
\end{figure}

\mbox{}\\[-8mm]\indent
The experimental measurement of $|\epsilon_K|=\left|A(K_L\to(\pi\pi)_{I=0})/A(K_S\to(\pi\pi)_{I=0})\right|$ together with 
a prediction of the kaon bag parameter
\begin{equation}\label{eq:BK}
	{ \hat B_K}=
        C(\mu)\frac{\langle \overline{{K}^0}|Q^{\Delta S=2}(\mu)| K^0\rangle}
                {\frac 83 f^2_K m_K^2}\,,
\end{equation}
define a hyperbola in the plane of the Wolfenstein parameters
$\hat \eta$ and $\hat \rho$ which constrains the apex of the
unitarity triangle shown in figure \ref{fig:CKM-triangle}.
Here $Q^{\Delta S=2}(\mu)$ equals the difference of the 
parity even and parity odd four quark operators
$O_{VV+AA}$ and $O_{VA+AV}$, respectively. 
Before presenting recent results for $\hat B_K$ I will now 
briefly discuss and compare how the major fermion 
discretisations fare in the computation of the matrix element in 
eqn.~(\ref{eq:BK}).

{\bf Wilson fermions}: As a result of the explicit breaking of chiral symmetry
due to the Wilson term
the $B_K$ operator mixes with four other operators $O_i$ 
\cite{Martinelli:1983ac},
\begin{equation}
O^{\Delta S=2}(\mu)=Z_{\rm VV+AA}(\mu,g_0^2)\left(
	O^{\rm latt}_{\rm VV+AA}+
          \sum\limits_{i=1}^4\Delta_i(g_0^2)O_i^{\rm latt}\right)\,,
\end{equation}
thus complicating the renormalisation process. 
It was realised in \cite{Becirevic:2000cy} that the parity
even operator $O_{VV+AA}$ is related to the parity odd operator
$O_{AV+VA}$ by an axial Ward identity and that the latter operator
  renormalises multiplicatively also for Wilson fermions. 
While the relevant matrix element
of the parity even operator can be determined from three-point functions, the
relevant matrix element for the parity odd operator has to be determined
from four point functions. In some sense one trades the uncertainties
due to the mixing in the case of the parity even operator for noisier signals
of the four point functions in the case of the parity odd operator.
Both approaches have been studied with dynamical fermions 
in \cite{Flynn:2004au,Mescia:2005ew}.

{\bf Twisted mass QCD}: 
The the parity even operator in QCD is related to the 
parity odd operator in tmQCD by an axial rotation 
\cite{Frezzotti:2000nk},
$\langle K^0|O_{\rm VV+AA}^{(0)}|\overline {K^0}\rangle_{\rm QCD}=-i
    \langle K^0|O_{\rm VA+AV}^{\,(\alpha)}|\overline {K^0}\rangle_{\rm tmQCD}\,.$
$B_K$ can now be computed from 3-pt. functions of $O_{\rm VA+AV}$ which
is expected to give a better signal than the 4-pt. function which
has to be computed in the case of standard Wilson fermions. 
A benchmark computation using this approach in quenched tmQCD
has recently been carried out by the ALPHA collaboration 
\cite{Dimopoulos:2006dm,Dimopoulos:2007cn}. In their simulation they used 
degenerate light and strange quark masses, thus neglecting possible 
$SU(3)$-breaking effects. The first case that was 
studied in that paper was with  a Wilson $s$-quark and two twisted light quarks
at twisting angle $\alpha=\pi/2$. Although non-degenerate $s$ and $d$ quarks
are in principle feasible in this approach, 
here the masses of the $s$ and the
$d$ quark were tuned to be degenerate, which is non-trivial since they have 
been discretised differently. 
The second case that ALPHA studied was $\alpha=\pi/4$ with twisted
$s$ and $d$ quarks in which case the quarks are automatically degenerate.
The simulations were carried out for a number of different 
lattice spacings, thus allowing for a very detailed
study of the approach to the continuum limit. The authors used the 
non-perturbatively  
computed (Schr\"odinger functional)
renormalisation constant for $O_{VA+AV}$ \cite{Guagnelli:2005zc}.
One further outcome of this
study is that the splitting of the meson spectrum due to the explicit
breaking of the flavour symmetry with twisted mass fermions reduces
as the continuum is approached where it is expected to vanish.

{\bf Staggered fermions}: Van de Water and Sharpe \cite{VandeWater:2005uq}
studied the transformation properties of the staggered $B_K$-operator
that couples to external kaons of taste $P$ which 
correspond to the lattice Goldstone kaon. The lattice representation of that
operator mixes with many other operators of all tastes. In current
simulations \cite{Gamiz:2006sq,Kim:2006ck} only the
operators with the same taste as the lattice Goldstone kaon are actually 
implemented. The corresponding mixing coefficients are computed in 
perturbation theory.
The mixing with other tastes at order $\alpha$ and higher orders in the strong 
coupling constant as well as 
lattice artefacts entering at order $a^2$ are described by staggered chiral
perturbation theory:
\begin{equation}
O_K^{\rm stagg, cont} = \underbrace{O_K^{\rm stagg}[{\rm taste\, P}] + 
               {\tiny \frac{\alpha}{4\pi}[{\rm taste\, P}]}}\limits_{
                \rm \tiny simulation}+
               \frac{\alpha}{4\pi}[{\rm wrong\, tastes}]+
        \underbrace{{\alpha^2}[{\rm all\, tastes}]}_{\rm unknown\,\, 2-loop}+
               \underbrace{a^2[{\rm all\, tastes}]}_{\rm discretisation}\,.
\end{equation}
The perturbative coefficients of the terms in $\alpha/(4\pi)$ are known. Since
the higher order perturbative coefficients are not known, they are counted
conservatively as $\alpha^2$ rather than $(\alpha/(4\pi))^2$. The above
expression has 37 free parameters which can be constrained e.g. by measuring 
the taste-splitting or by first determining a sub-set of parameters at 
a single lattice spacing.
It is also known that HYP-smearing \cite{Hasenfratz:2001hp}
reduces the mixing and taste 
breaking and also improves the convergence of perturbation theory and may
thus yield a more favourable power counting \cite{Lee:2003sk,Lee:2006cm}.
It has to be mentioned that non-perturbative renormalisation is in principle 
possible. However, all current simulations of $B_K$ with staggered fermions
rely on perturbative renormalization.

{\bf Domain wall fermions and overlap fermions}:
With chiral fermion formulations the parity-even operator $O_{VV+AA}$ 
renormalises multiplicatively. For domain wall fermions residual mixing with
wrong chirality operators was discussed in detail in \cite{Aoki:2005ga}. It is 
suppressed by $(am_{\rm res})^2$ and therefore negligible 
 (e.g. $am_{\rm res}\approx 10^{-3}$ for the current
RBC+UKQCD domain wall fermion data set \cite{Peter}). 
The chiral symmetry of these actions provides automatic
$O(a)$-improvement and continuum chiral perturbation theory
can be used.

{\bf Recent simulations}:
Current efforts for the calculation of $B_K$ on the lattice with dynamical
fermions are summarised
in table \ref{tab:BK_params}. HPQCD+UKQCD \cite{Gamiz:2006sq} 
on the one hand and
Bae, Kim, Lee and Sharpe \cite{Kim:2006ck}
on the other hand are using 2+1 flavour staggered
quarks (MILC configurations \cite{MILCconfigs})
at the same simulation parameters. Both collaborations use HYP smeared 
\cite{Hasenfratz:2001hp}
valence quarks, thus simulating a partially quenched theory.
While HPQCD+UKQCD is 
simulating for degenerate $s$- and $d$-quarks only, Bae and collaborators
also investigate $SU(3)$-breaking effects and fit their
data to continuum partially quenched 
chiral perturbation theory \cite{VandeWater:2005uq}. 
It remains to be studied in 
detail whether taste breaking effects for HYP-valence quarks are sufficiently
suppressed that one can apply continuum chiral perturbation theory
instead of staggered chiral perturbation theory in 
order to reliably describe and extrapolate the data. In order to estimate 
cut-off effects, HPQCD+UKQCD compare to simulations at various
lattice spacings but otherwise similar
simulation parameters within the quenched approximation. The underlying 
assumption is that cut-off effects in the quenched and unquenched theory
behave similarly. 
\begin{table}
\small
 \begin{center}
 \begin{tabular}{l@{\hspace{-4mm}}c@{\hspace{-1mm}}c@{\hspace{-1mm}}c@{\hspace{-2mm}}c@{\hspace{1mm}}c@{\hspace{1mm}}c@{\hspace{1mm}}c@{\hspace{4mm}}c@{\hspace{2mm}}c@{\hspace{0mm}}c@{\hspace{0mm}}lc}
 \hline\\[1.5mm]
                &&$N_f$&&a&$m_\pi L$
                &$m_\pi$/MeV$^\dagger$
                &\begin{rotate}{40}{\tiny renorm.}\end{rotate}
                &\begin{rotate}{40}{\tiny degenerate}\end{rotate}
                &\begin{rotate}{40}{\tiny non-degenerate}\end{rotate}
                &$\chi$PT
		&\multicolumn{1}{c}{$\hat B_K$\,\,\,\,}\\[0mm]
 \hline\hline\\[-5mm]
 {\hspace{-2.0mm}\begin{tabular}{l}HPQCD+\\[-1.5mm]UKQCD\end{tabular}}&{ \cite{Gamiz:2006sq}} &2+1& $\rm KS_{\rm MILC}^{\rm HYP}$ &
                \begin{tabular}{c}
                        0.125\\[-3.0mm]
                        {\tiny quenched}\\[-3.5mm]
                        {\tiny scaling}
                \end{tabular} & 4.5&360&PT&\textbullet& &-&$0.83(18)$\\[-1mm]
 {\hspace{-2mm}\begin{tabular}{l}Bae, Kim,\\[-2mm]Lee, Sharpe\end{tabular} }&
	{ \cite{Kim:2006ck}}&
	{2+1}& 
	{$\rm KS_{\rm MILC}^{\rm HYP}$} &
        {
	      \begin{tabular}{c}
                        0.125
               \end{tabular}} & 
	{4.5}&
	{360}&
	{PT}&
	{\textbullet}&
	{\textbullet}&
	{NLO}
	\\[-1mm]
 {\hspace{-2.0mm}\begin{tabular}{l}RBC+\\[-2mm]UKQCD\end{tabular}}&{
	\begin{tabular}{c}
	 \cite{Antonio:2007pb}\\[-1mm]
	\cite{Antonio,Cohen}
	\end{tabular}}   &2+1& $\rm DWF$ &
                \begin{tabular}{c}
                        0.11\\[-3.0mm]
                        {\tiny quenched}\\[-3.5mm]
                        {\tiny scaling}
                \end{tabular} & 4.6 &330&NPR&\textbullet&\textbullet&\hspace{-1mm}NLO&0.720(39)\\[-2mm]
 {JLQCD}&{ \cite{Yamada}}       &2&
                \begin{tabular}{c}
                Overlap\\[-2.0mm]
                fix. top.
                \end{tabular}&0.12
                & 2.7 &290&NPR&\textbullet&\textbullet&
                \begin{tabular}{c}
                        NLO\\
                \end{tabular}&$0.723(12)^\ddagger$
                \\[-0mm]
 \hline\\[-6mm]
	\multicolumn{12}{c}{ {\tiny $\dagger$ for lightest pion; $^\ddagger$ statistical error only}}\\
 \hline\hline\\
 \end{tabular}\\
 \end{center}\mbox{}\\[-15mm]
 \caption{Summary of current $B_K$-calculations.}
 \label{tab:BK_params}
\end{table}
RBC+UKQCD \cite{Antonio:2007pb,Antonio,Peter}
and JLQCD \cite{Yamada} use chiral fermions with 2+1 flavours of
domain wall quarks and 2 flavours of overlap quarks, respectively. RBC+UKQCD
estimate the cut-off effects from the experience with the  quenched 
case \cite{Aoki:2005ga,AliKhan:2001wr} and both collaborations consider
non-degenerate $s$- and $d$-quarks in the partially quenched framework and
have renormalised the $B_K$ operator non-perturbatively in the RI-MOM scheme
\cite{Martinelli:1994ty}.
JLQCD's overlap quarks
are simulated at fixed topological charge and the corresponding
finite volume effects are estimated to be at the percent level; this estimate
is obtained by comparing results from
different charge sectors at constant volume and quark mass.
JLQCD has not yet finalised the error analysis for their value of $B_K$ and
it will be interesting to see the the full impact of finite volume effects 
($m_\pi L\approx 2.7$) on the final error budget.
Since the results of HPQCD+UKQCD and Bae, Kim, Lee and Sharpe have been nicely
discussed in last year's plenary talk \cite{Lee:2003sk} I concentrate here 
on the ones by RBC+UKQCD and JLQCD.

\begin{figure}
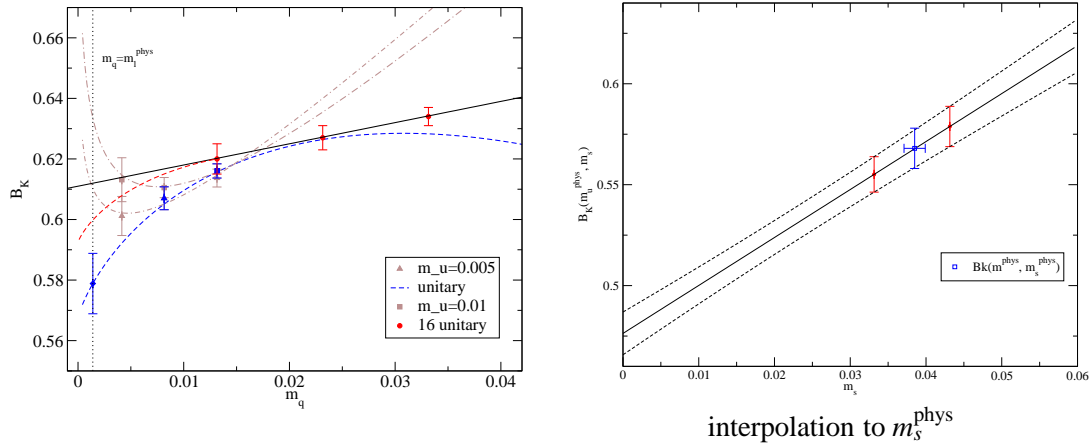

        \begin{minipage}[c]{.49\linewidth}
        \begin{center}
         \epsfig{scale=.28,file=./bildla/bk24_16b.eps}\\
        \end{center}
        \end{minipage}
        \begin{minipage}[c]{.49\linewidth}
        \mbox{}\\[2mm]
        \begin{center}
         \epsfig{scale=.28,file=./bildla/msExtrap-2.eps}\\
        interpolation to $m_s^{\rm phys}$\\[2.5mm]
        \mbox{}\\
        \end{center}
        \end{minipage}\\[-10mm]
\caption{RBC+UKQCD results for $B_K$ 
	\cite{Antonio,Antonio:2007pb}.}\label{fig:BK_UKQCD}
\end{figure}

RBC+UKQCD simulated on two volumes ($L=1.8$fm and $L=2.7$fm) 
and as can be gathered from
the l.h.s. plot in figure \ref{fig:BK_UKQCD}, no significant finite volume 
effects were seen. The plot shows a number of partially quenched data points
for configurations with sea quark masses that correspond to 
approximately 330 MeV and 420 MeV pions. The data was described and extrapolated
using NLO $SU(2)\times SU(2)$ chiral perturbation
theory \cite{Sharpe:1995qp} at fixed values of the strange quark mass
which was then subsequently interpolated to the physical point (r.h.s. plot
in figure \ref{fig:BK_UKQCD}).
\begin{figure}
 \hspace{-5mm}\begin{minipage}{.49\linewidth}
 \begin{center}
 {\epsfig{scale=.4,file=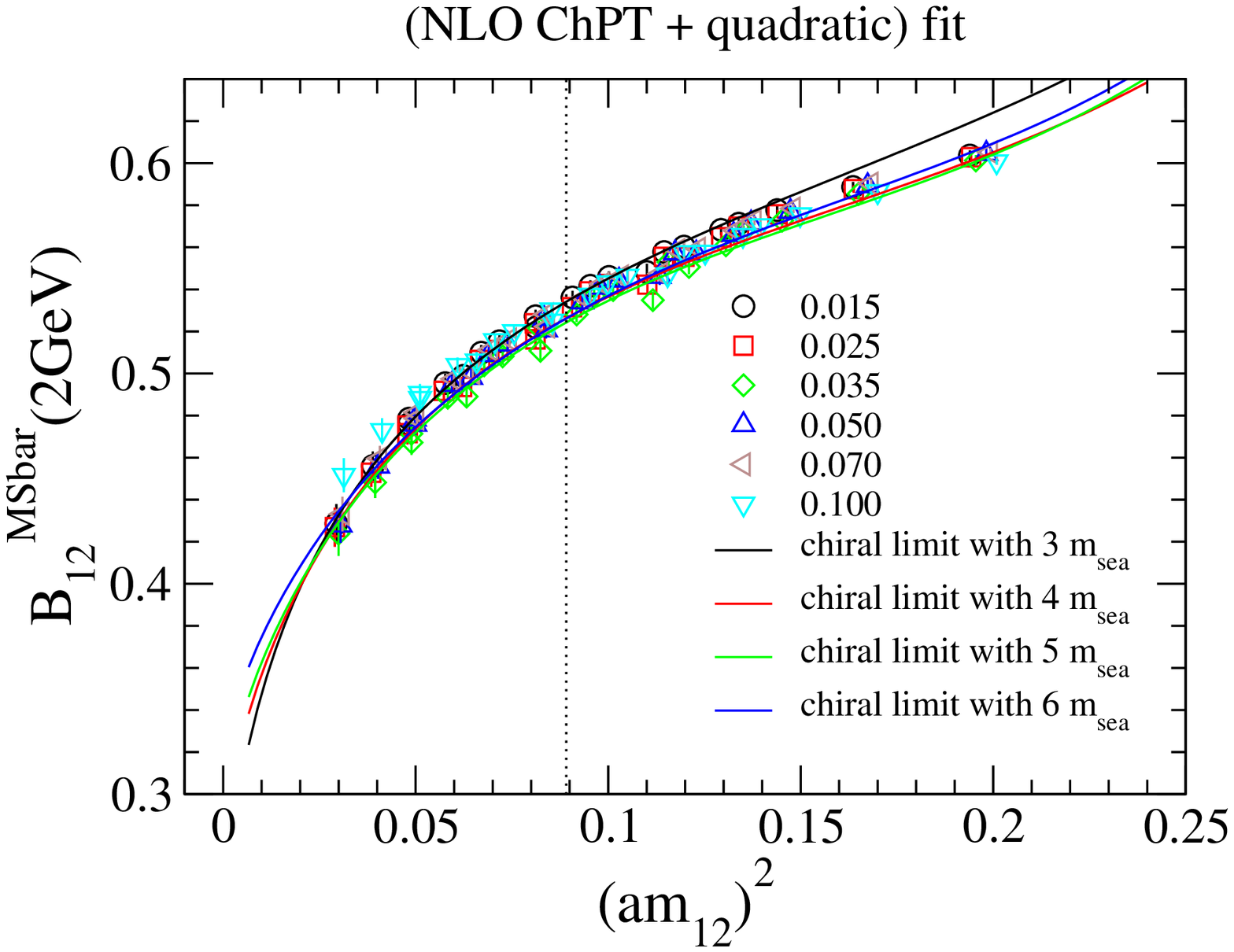}}\\[-2mm]
        \hspace*{12mm}full data set
 \end{center}
 \end{minipage}
 \begin{minipage}{.49\linewidth}
 \begin{center}
 {\epsfig{scale=.4,file=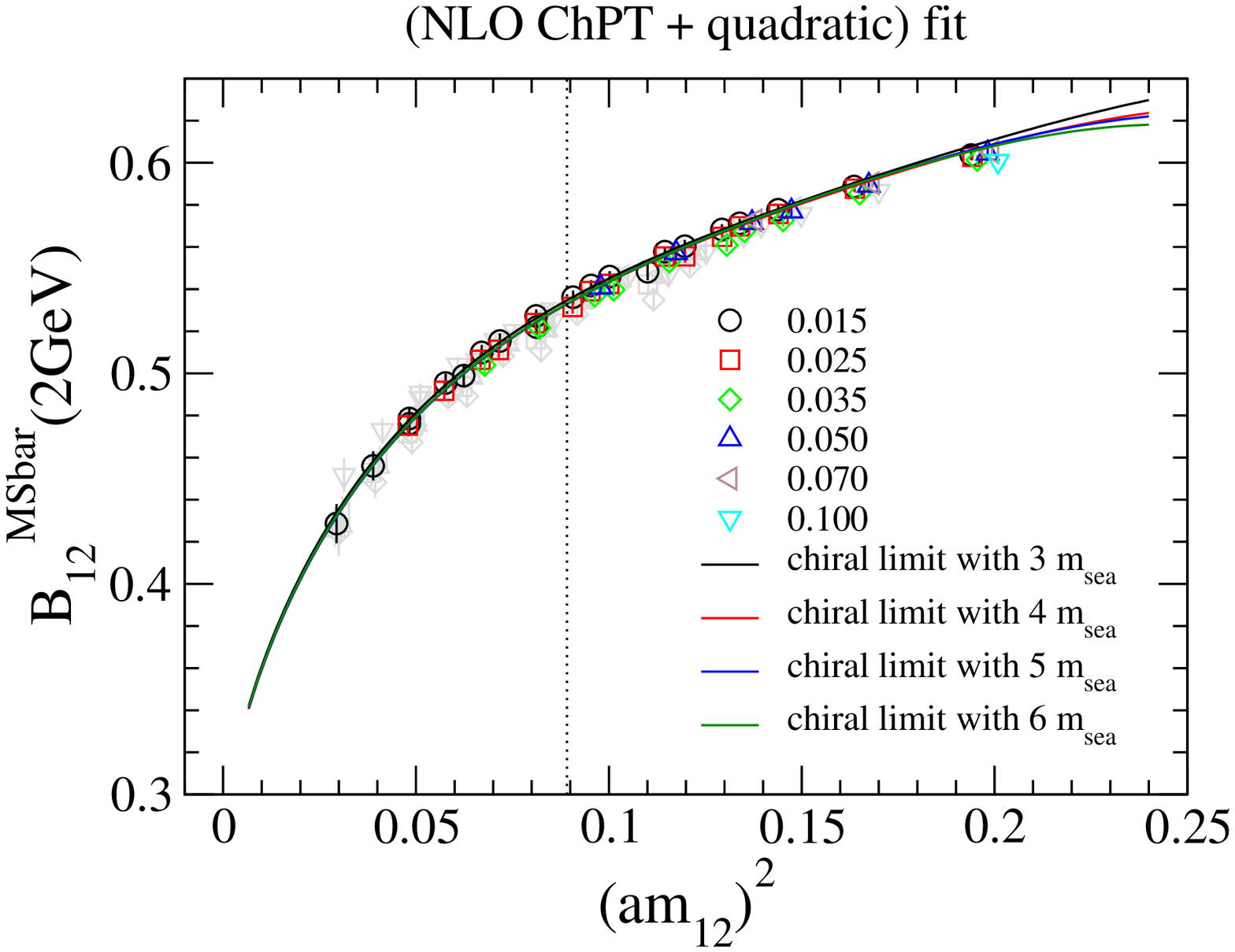}}\\
        \hspace*{10mm}reduced data set
        \mbox{}\\
 \end{center}
 \end{minipage}
\caption{JLQCD results for $B_K$ \cite{Yamada}.}\label{fig:BK_JLQCD}
\end{figure}
JLQCD's results are illustrated in figure \ref{fig:BK_JLQCD}. They 
used NLO $SU(3)\times SU(3)$ partially quenched chiral 
perturbation theory \cite{VandeWater:2005uq} to extrapolate their
data to the physical point. The l.h.s. plot in figure \ref{fig:BK_JLQCD}
shows a significant dependence of the fit-results on the number of 
data points that were included into the fit. 
After applying the cut $m_l\le m_s/2$ their fits however
turned out to be stable and as the r.h.s. plot of figure \ref{fig:BK_JLQCD}
suggests that the fits all agree after inclusion of  
an additional NNLO analytic term into their fit-ansatz. 

\begin{figure}
 \begin{center}
 \psfrag{hatBK}[t][t][1][0]{${\hat B}_K$}
\epsfig{scale=.25,angle=-90,file=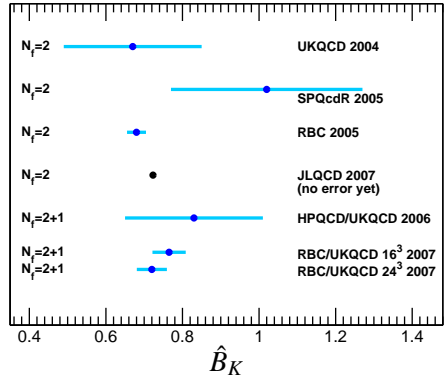}
 \end{center}\mbox{}\\[-10mm]
 \caption{Summary of recent lattice simulations with dynamical fermions; 
	for the JLQCD-result only the central value is shown since
	the error analysis is not yet finished.} 
\label{fig:BK_results}
\end{figure}

A summary of all recent lattice computations of $B_K$ with dynamical
fermions is given in figure \ref{fig:BK_results} and I quote the  
numerical values of the most advanced simulations
in table \ref{tab:BK_params}. The available results
for the computation of the $B_K$-operator $O_{VA+AV}$
with Wilson fermions 
either directly or via the axial Ward identity method have rather
large errors. The reason is the lack of $O(a)$-improvement of the 
$B_K$-operator, rather
heavy light quark masses and perturbative renormalization in these simulations.
Also the HPQCD+UKQCD result from staggered fermions
has a large error bar which contains a large contribution
from the perturbative
treatment of the operator mixing. The implementation of non-perturbative 
renormalisation for the staggered $B_K$-operators should be considered 
in the future. 
The results with the
smallest error bars have all been computed using chiral fermions
thanks to the multiplicative
renormalisation of $O_{AA+VV}$ which was realized non-perturbatively in both
simulations. JLQCD currently only
quotes their central value for $B_K$ since the error analysis is not yet 
finalized.
The current best estimates for  $\hat{B}_K$ from lattice QCD is the one
by RBC+UKQCD \cite{Antonio},
\begin{equation}
\begin{array}{lcrcl}
\hat{B}_K &=& 0.720(39)\,.	\\[1mm]
\end{array}
\end{equation}
This result is compatible with the result of the previous $N_f=2$-simulations
with domain wall fermions by RBC \cite{Aoki:2004ht} and also with 
the central value of the new simulation by JLQCD \cite{Yamada}.

{\bf New developments}: Aubin, Laiho and Van de Water 
\cite{Aubin:2006hg} developed the partially quenched chiral perturbation
theory for domain wall valence fermions combined with 
AsqTad staggered sea quarks. 
In this mixed action ansatz \cite{Bar:2005tu} the symmetry properties
of the domain wall valence quarks protects the $B_K$-operator from 
mixing with operators of non-trivial taste structure. Compared to 
continuum partially quenched chiral perturbation theory 
\cite{VandeWater:2005uq} there are only two additional parameters. With
the large set of staggered sea quark configurations by the MILC
collaboration \cite{Bernard} a computation of $B_K$ using this approach
seems feasible and very preliminary results have been presented
at this conference \cite{Aubin_firstBK}. 

The scale evolution of the $B_K$ operator is usually carried out in 
perturbation theory. In order to remove the uncertainty due to
perturbation theory the ALPHA collaboration has determined the
scale evolution of the parity-odd operator $O_{VA+AV}$ non-perturbatively 
in the continuum limit of the 
$N_f=0,2$ Schr\"odinger functional scheme \cite{Pena,Dimopoulos:2006es}.
If $O_{VA+AV}$ is also renormalised in the 
Schr\"odinger functional scheme, the discretisation independent result for
the scale evolution can be used to 
compute the running of $B_K$ as obtained with Wilson fermions using
the axial Ward identity method and the running of $B_K$ as obtained using 
twisted mass QCD or Ginsparg-Wilson-type fermions. 

Assuming that $O_{VA+AV}$ in the continuum limit of twisted mass QCD
has been non-perturbatively renormalised at the scale $\mu$, it was
suggested in \cite{Dimopoulos:2006ma} that the renormalisation constant for 
the corresponding operator determined using Ginsparg-Wilson-type fermions,
$Z^{GW}(\mu,g_0)$, at non-vanishing lattice spacing could be defined 
via 
\begin{equation}
{O_{VA+AV}(\mu,m_{PS})}|_{\rm c.l.\, of\, tmQCD}
	= Z^{\rm GW}(\mu,g_0) O_{VA+AV}^{\rm GW}(g_0,m_{PS}) + O(a^2)\,,
\end{equation}
once $m_{PS}$ on the l.h.s. and r.h.s. have been tuned to the same
value. The identity holds due to the automatic $O(a)$ improvement of chiral
fermions.
\subsection{\textit{The Holy Grail} - direct $CP$-violation}\mbox{}\\[-8mm]\indent
$CP$-violation is described in terms of the isospin amplitudes
\begin{equation}
\small
        A(K^0\to \pi^+\pi^-)=\sqrt{\frac 23}A_0e^{i\delta_0}+
                               \sqrt{\frac13} A_2e^{i\delta_2}\;\;{\rm and}\;\;
        A(K^0\to \pi^0\pi^0)=\sqrt{\frac 23}A_0e^{i\delta_0}-
                                \sqrt{\frac13} A_2e^{i\delta_2}\\
\end{equation}
from which one constructs the parameters 
$\epsilon^\prime = \frac{\omega}{\sqrt 2}e^{i\phi}\left(
         \frac{{\rm Im} A_2}{{\rm Re} A_2}-\frac{{\rm Im} A_0}{{\rm Re} A_0}
                \right)$ and 
$\omega = \frac{{\rm Re} A_2}{{\rm Re} A_0}$ describing direct
$CP$-violation and the $\Delta I=1/2$-rule, respectively.
The isospin amplitudes are defined in terms of the matrix element
\begin{equation}
\langle \pi\pi(I)|-i{\cal H}|K^0\rangle=A_Ie^{i\delta_I}\,.
\end{equation}
where the  relevant effective Hamiltonians \cite{Georgi,Buchalla:1995vs}
are 
\begin{equation}
 \begin{array}{llll}
 {\rm for\;} {u,d,c,s}&{{\cal H}^{\Delta S=1}_c}=\frac{G_F}{\sqrt{2}}V_{ud}V_{us}^\ast
        \sum\limits_{\sigma=\pm}\big\{
        k_1^\sigma(\mu){\cal O}_1^\sigma(\mu)+
        k_2^\sigma(\mu){\cal O}_2^\sigma(\mu)
        \big\}\,,&\\[3mm]
 {\rm for\;}u,d,s&{{\cal H}^{\Delta S=1}}=\frac{G_F}{\sqrt{2}}V_{ud}V_{us}^\ast
        \sum\limits_{i=1}^{10}C_i(\mu)O_i(\mu)\,,
        &
 \end{array}
\end{equation}\\[-4mm]
for the four- and three flavour case, respectively, where
${\cal O}^\pm_{1,2}$ and ${ O}_{1,2}$ are current-current operators,
$O_{3,4,5,6}$ are QCD penguin operators and $O_{7,8,9,10}$ are EW 
penguin operators. 
${\cal H}^{\Delta S=1}_c$ does not contain penguin diagrams. 
Giusti et al. \cite{Giusti:2004an} are
studying this Hamiltonian in order to qualitatively
 assess the role of the charm 
quark in the $\Delta I=1/2$ rule \cite{Giusti:2004an} and their programme has
been nicely reported in Hernandez's plenary talk at Lattice 2006 
\cite{Hernandez:2006au}. 
Large scale dynamical lattice simulations with chiral fermions with 
the aim to compute the physical values of 
$|\epsilon^\prime/\epsilon|$ 
and $\omega$ use ${{\cal H}^{\Delta S=1}}$ where the
physical charm quark that is too heavy for current lattice simulations has
been integrated out.
Two-pion final states are notoriously difficult to handle on the lattice
\cite{Maiani:1990ca}. Instead of computing 
$\langle \pi\pi(I)|-i{\cal H}|K^0\rangle$  directly 
one uses chiral perturbation theory at LO and NLO
in order to relate the matrix elements of interest
to $K\to{\rm vacuum}$, $K\to\pi$ and $K\to\pi\pi$ matrix elements at 
unphysical kinematics which can be computed on the lattice more easily
\cite{Bernard:1985wf,Laiho:2002jq,Lin:2002nq,Laiho:2003uy}. In two 
impressive works by the CP-PACS \cite{Noaki:2001un} and the RBC 
collaboration \cite{Blum:2001xb} the
approach has been demonstrated to work. 
The use of the quenched approximation
and the rather large values of the light quark masses in these
calculations  of course have to be overcome.  To this end 
Bae, Kim and Lee carried out exploratory studies of the 
calculation of $O_{7,8}^{(3/2)}$ using $N_f=2+1$ flavours of 
staggered quarks \cite{Bae:2005su,Kim:2005st}. Very recently
the RBC-collaboration has started to repeat their calculation but
this time with $N_f=2+1$ flavours of dynamical domain wall fermions.
Their lattice spacing is $a^{-1}=1.73$GeV and the spatial size of the lattice
is $L=2.7$fm. The collaboration is planning to simulate for 2
sea quark masses corresponding to pion masses of $m_\pi\approx 330$MeV
and $410$MeV, combined with 6 partially quenched valence quark simulation 
points and a fixed strange quark mass. 
All operators will be renormalised non-perturbatively (RI/MOM scheme 
\cite{Martinelli:1994ty}).
One of the many technical difficulties 
that arise in the project is the appearance of power
divergencies which have to be subtracted like e.g.
\begin{equation}
\langle \pi^+|O_i^{(1/2)}|K^+\rangle
                +{\eta_{1,i}(m_s-m_d)\langle 0 |(\bar s d)|K^+\rangle}\,,
\end{equation}
where the subtraction coefficient $\eta_{1,i}$ is determined from the
slope of the quark mass dependence of the ratio of matrix elements
\begin{equation}\label{eq:DS1_sub}
\frac{\langle 0 |O_i|K^0\rangle}
                {\langle 0|(\bar s\gamma_5d)|K^0\rangle}=
                \eta_{0,i}+{ \eta_{1,i}}(m_s-m_d)\,.
\end{equation}
\begin{figure}
\begin{center}
\epsfig{scale=.22,file=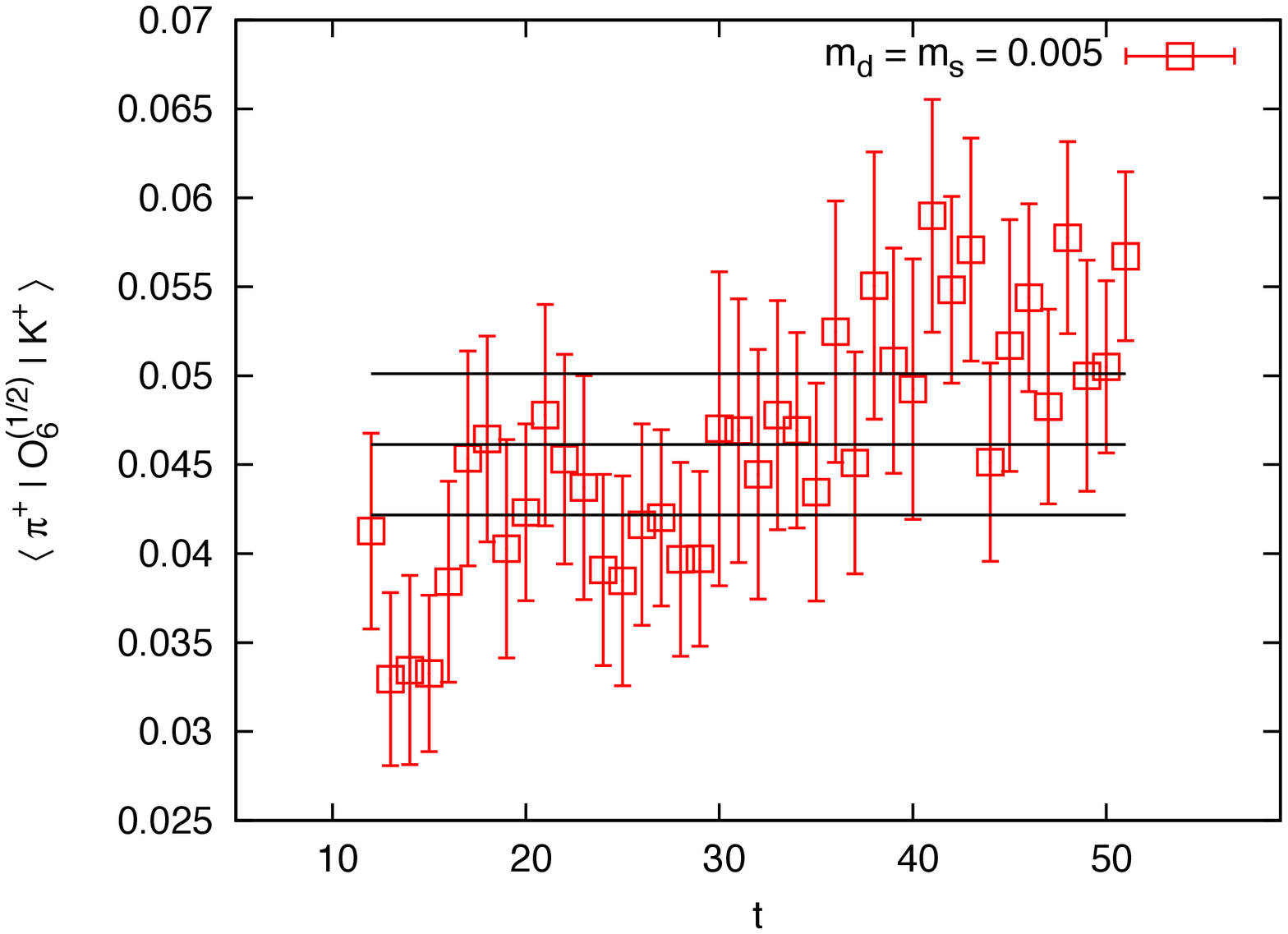}
         \epsfig{scale=.22,file=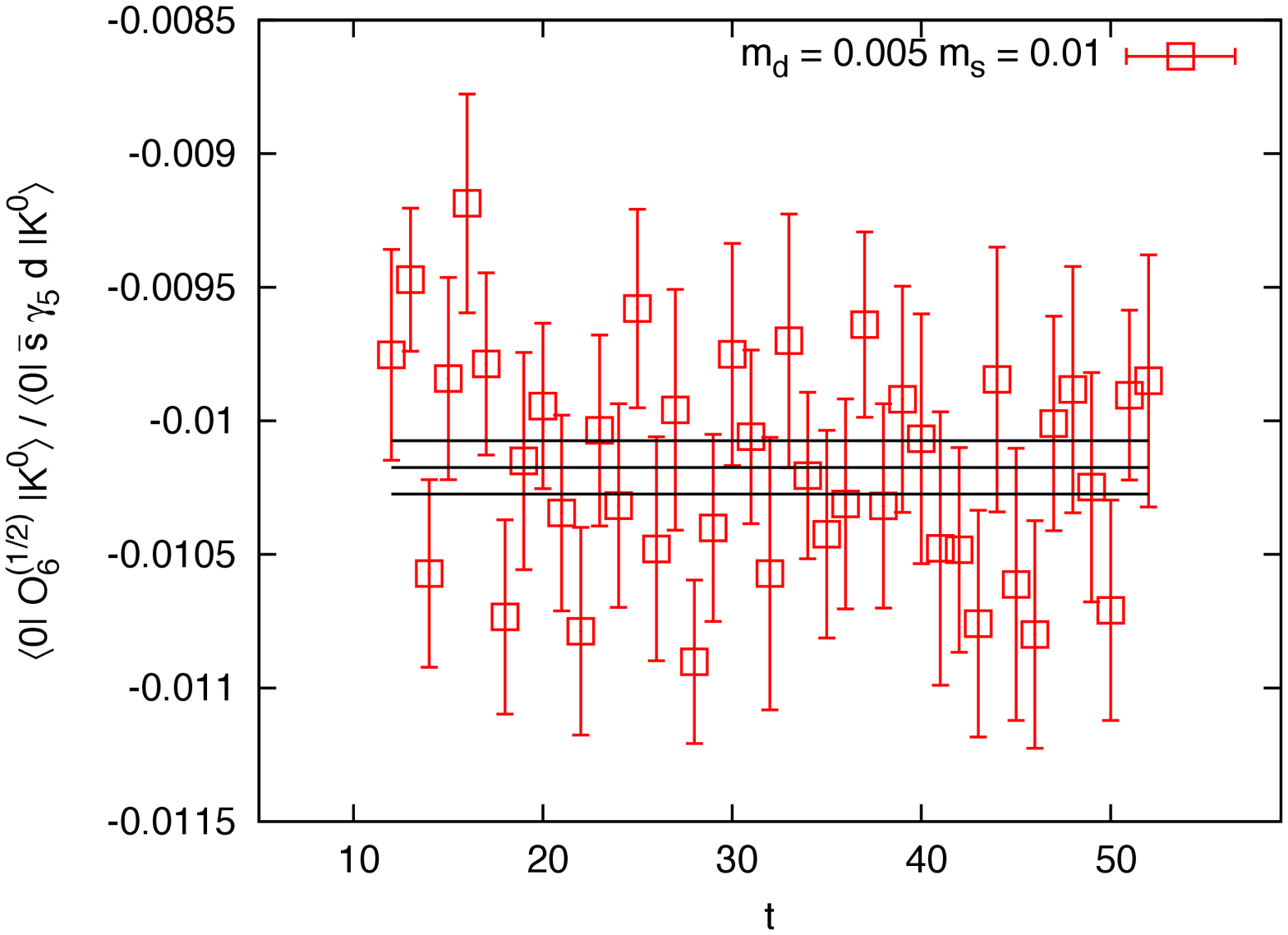}
         \epsfig{scale=.22,file=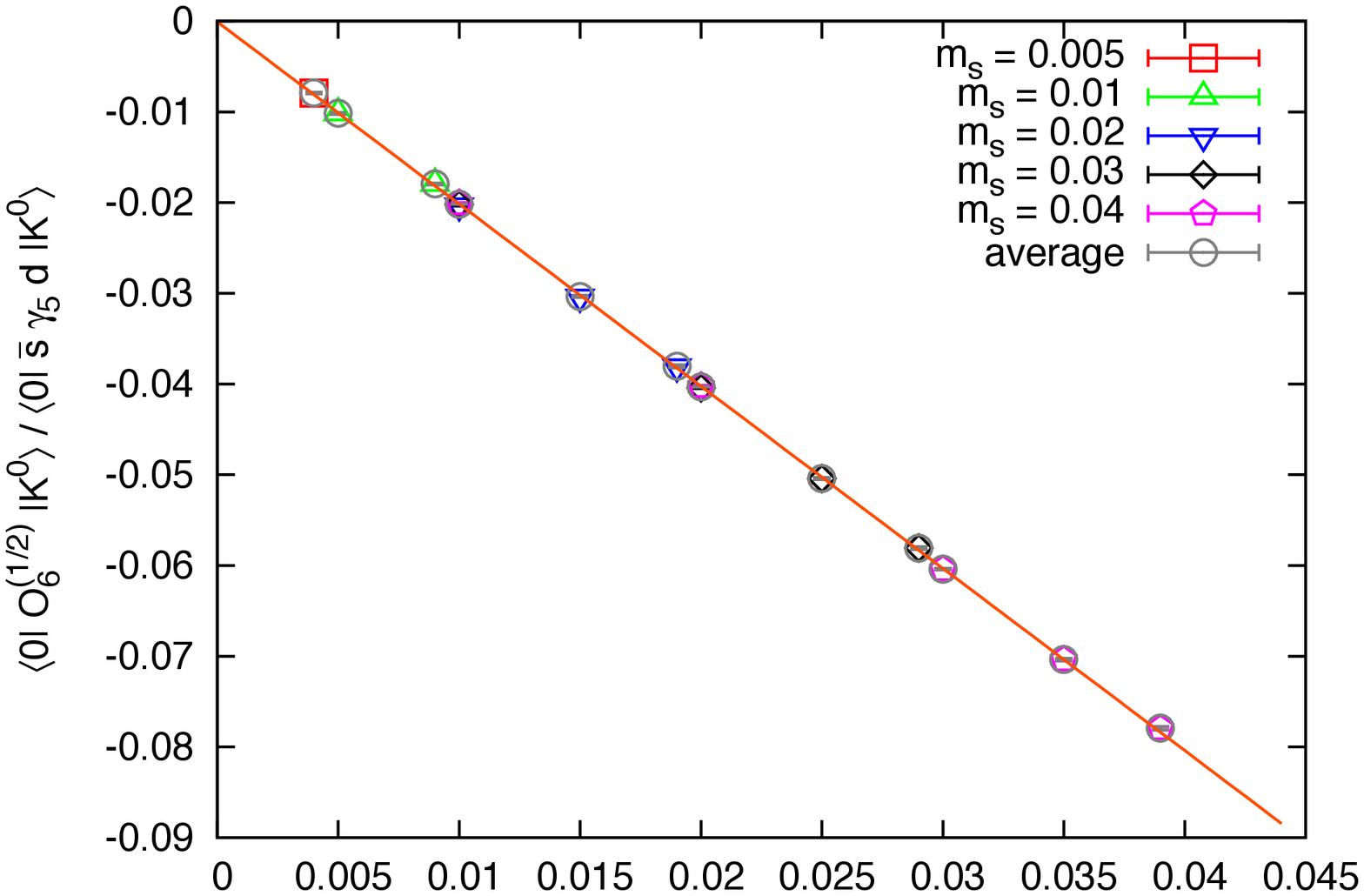}\\
\epsfig{scale=.24,file=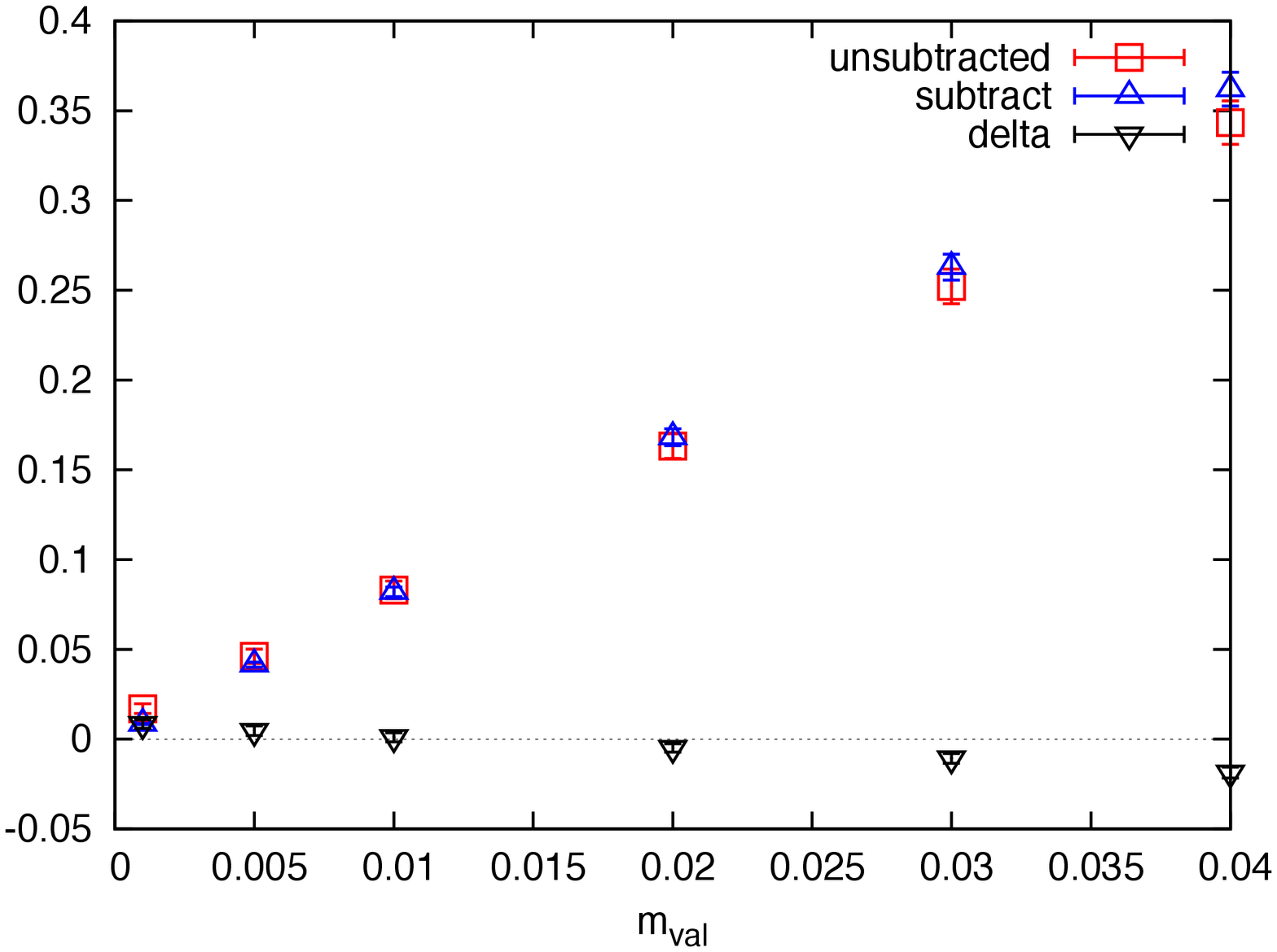}
\end{center}\mbox{}\\[-14mm]
\caption{Power subtractions in the computation of 
	\protect$\Delta S=1$ matrix elements:
	First row: Signal for \protect$\langle \pi^+|O_i^{(1/2)}|K^+\rangle$
	and \protect$\frac{\langle 0 |O_i|K^0\rangle}
                {\langle 0|(\bar s\gamma_5d)|K^0\rangle}$
	and the mass dependence of the latter. 
	Second row: The power subtraction as in equation
	(\protect\ref{eq:DS1_sub}).}\label{fig:DS1_power}
\end{figure}
The results for the 
involved matrix elements and the mass dependence of the ratio 
are illustrated in the plots in the first line of figure \ref{fig:DS1_power}. 
That the subtraction (\ref{eq:DS1_sub}) works very nicely in practice
is illustrated in the second line of figure \ref{fig:DS1_power}. The
residual chiral symmetry breaking of domain wall fermions leads to 
an ambiguity in the normalisation of the subtraction procedure. This is 
however irrelevant here since only the slope of the subtracted matrix element 
with respect to the quark masses is of interest.
The example given here is representative for the technical challenges 
encountered in this project which the RBC-collaboration is now carrying
over to the NLO-level of chiral perturbation theory.

For completeness I also want to mention the study in \cite{Boucaud:2004aa}
where the $K\to\pi\pi$ matrix element for $O_{7,8}^{(3/2)}$ was evaluated
directly for unphysical kinematics and also another exploratory study of 
$K\to\pi\pi$ at NLO using $N_f=2$ flavours of domain wall fermions in 
\cite{Noaki:2005zw}.

Another technical difficulty in the calculation of $K\to\pi\pi$ 
processes in the $\Delta I=1/2$-channel
is the computation of quark-disconnected diagrams that appear
as a result of Wick contractions. Kim and Sachrajda \cite{Kim} propose to 
compute the process $K\pi^-\to\pi^-$ instead of  $K\to\pi^+\pi^-$. It turns
out that the time-like disconnectedness 
in the latter process is replaced by a space-like disconnectedness in 
the former which is technically much easier to handle. The relation 
between the two processes has been worked out in 
NLO chiral perturbation theory. \\[-5mm]
\section{Summary}\mbox{}\\[-8mm]\indent
In this talk I have reviewed recent progress in kaon physics. 
I discussed the determination of $|V_{us}|$ from lattice predictions for
$f_K/f_\pi$ and for $f_+^{K\pi}(0)$. The computation of the latter
with a precision of $0.5\%$ in full QCD is now feasible. The use
of all-to-all propagators in the calculation of the relevant meson
correlation functions and a  new approach using partially 
twisted boundary conditions will very likely further reduce the error 
in the near future - new calculations based on these techniques are under
way.  
The determination of $|V_{us}|$ from $f_K/f_\pi$ will  become
an independent competitor for the determination from $f_+^{K\pi}(0)$. I expect
the current precision of the various
 results  (the best being 
0.6\% in the case of HPQCD/UKQCD) to be updated in the near
future as collaborations continue extending their sets of gauge field
configurations towards lighter masses and also towards smaller lattice spacings.
This will then also allow to further improve the  quality of the 
combined analysis of $|V_{us}|$ from $f_K/f_\pi$ and from $f_+^{K\pi}(0)$ in
view of CKM-unitarity  which will be exciting to monitor.
 
For $B_K$ the results from the various collaborations agree within errors. The
magnitude of the errors however varies strongly and the results from
chiral fermion formulations currently look most promising with an error
of about 7\% in full QCD. Collaborations using staggered quarks in their
calculation of $B_K$ should implement non-perturbative renormalisation in order to 
reduce the currently rather large error bars.

For $\Delta S=1$ the RBC collaboration is in the middle of a large scale
project with light domain wall fermions in large volume and it will be 
interesting to see whether they can reproduce the measured value
of $|\epsilon^\prime/\epsilon|$ and shed light on the $\Delta I=1/2$-rule - 
the simulations are however technically extremely demanding.\\[0mm]

{\bf Acknowledgements:} I want to thank 
M.~Antonelli,
D.~Antonio,
C.~Bernard,
P.~Boyle,
C.~Davies,
J.~Flynn,
C.~Kim,
F.~Mescia,
M.~Moulson,
C.~Sachrajda,
G.~Schierholz,
E.~Scholz,
C.~Tarantino,
H.~Wittig and
J.~Zanotti
for their kind help during the preparation of the talk and the proceeding. 
This work was 
supported by PPA/G/S/2002/00467 and PPA/G/O/2002/0046.\\[-7mm]
\bibliographystyle{h-elsevier2}
\bibliography{juettner_lat07}

\begin{thebibliography}{100}

\bibitem{Peter}
P. Boyle, 2+1 flavour {D}omain {W}all fermion simulations by the {RBC} and
  {UKQCD} collaborations,
\newblock (2007), arXiv:0710.5880 [hep-lat].

\bibitem{Matsufuru}
H. Matsufuru, f.t. JLQCD and . TWCD-Collaborations, Exploring the chiral regime
  with dynamical overlap fermions,
\newblock (2007), arXiv:0710.4225 [hep-lat].

\bibitem{Urbach}
C. Urbach, Lattice {QCD} with two light wilson quarks and maximally twisted
  mass,
\newblock PoS LAT2007 (2007) 022, arXiv:0710.1517 [hep-lat].

\bibitem{Kuramashi}
Y. Kuramashi, Dynamical wilson quark simulations toward the physical point,
\newblock PoS LAT2007 (2007) 017.

\bibitem{Creutz}
M. Creutz, Why rooting fails,
\newblock (2007), arXiv:0708.1295 [hep-lat].

\bibitem{Kronfeld}
A.S. Kronfeld, Lattice gauge theory with staggered fermions: how, where, and
  why (not),
\newblock (2007), arXiv:0711.0699 [hep-lat].

\bibitem{Sharpe:2006re}
S.R. Sharpe, Rooted staggered fermions: Good, bad or ugly?,
\newblock PoS LAT2006 (2006) 022, hep-lat/0610094.

\bibitem{Cabibbo:1963yz}
N. Cabibbo, Unitary symmetry and leptonic decays,
\newblock Phys. Rev. Lett. 10 (1963) 531.

\bibitem{Kobayashi:1973fv}
M. Kobayashi and T. Maskawa, {CP} violation in the renormalizable theory of
  weak interaction,
\newblock Prog. Theor. Phys. 49 (1973) 652.

\bibitem{Marciano:2004uf}
W.J. Marciano, Precise determination of {$|V_{us}|$} from lattice calculations
  of pseudoscalar decay constants,
\newblock Phys. Rev. Lett. 93 (2004) 231803, hep-ph/0402299.

\bibitem{PDBook}
W.M. {Yao} et~al., {Review of Particle Physics},
\newblock {Journal of Physics G} 33 (2006) 1+.

\bibitem{Marciano_Kaon}
W. Marciano, Implications of {CKM} unitariy,
\newblock  Kaon 2007 (2007) 003.

\bibitem{Marciano:2005ec}
W.J. Marciano and A. Sirlin, Improved calculation of electroweak radiative
  corrections and the value of {$V_{ud}$},
\newblock Phys. Rev. Lett. 96 (2006) 032002, hep-ph/0510099.

\bibitem{Schierholz}
G. Schierholz, Probing the chiral limit with clover fermions {I}: {T}he meson
  sector,
\newblock PoS LAT2007 (2007) 133.

\bibitem{Tarantino}
B. Blossier et~al., Light quark masses and pseudoscalar decay constants from
  {$N_f=2$} lattice {QCD} with twisted mass fermions,
\newblock (2007), arXiv:0709.4574 [hep-lat].

\bibitem{ETM:2007}
ETM, B. Blossier et~al., Light quark masses and pseudoscalar decay constants
  from {$N_f=2$} lattice {QCD} with twisted mass fermions,
\newblock (2007), arXiv:0709.4574 [hep-lat].

\bibitem{Aubin:2004fs}
MILC, C. Aubin et~al., Light pseudoscalar decay constants, quark masses, and
  low energy constants from three-flavor lattice {QCD},
\newblock Phys. Rev. D70 (2004) 114501, hep-lat/0407028.

\bibitem{Bernard:2006wx}
MILC, C. Bernard et~al., Update on the physics of light pseudoscalar mesons,
\newblock PoS LAT2006 (2006) 163, hep-lat/0609053.

\bibitem{Bernard}
C. Bernard et~al., Status of the {MILC} light pseudoscalar meson project,
\newblock (2007), arXiv:0710.1118 [hep-lat].

\bibitem{Follana:2007uv}
HPQCD, E. Follana et~al., High precision determination of the {$\pi$}, {$K$},
  {$D$} and {$D_s$} decay constants from lattice {QCD},
\newblock (2007), arXiv:0706.1726 [hep-lat].

\bibitem{Beane:2006kx}
S.R. Beane et~al., {$f_K/f_\pi$} in full {QCD} with domain wall valence quarks,
\newblock Phys. Rev. D75 (2007) 094501, hep-lat/0606023.

\bibitem{Allton:2007hx}
RBC and UKQCD, C. Allton et~al., 2+1 flavor domain wall {QCD} on a {$(2{\rm
  fm})^3$} lattice: {L}ight meson spectroscopy with {$L_s=16$},
\newblock Phys. Rev. D76 (2007) 014504, hep-lat/0701013.

\bibitem{Scholz}
UKQCD, M. Lin and E.E. Scholz, Chiral {L}imit and {L}ight {Q}uark {M}asses in
  2+1 {F}lavor {D}omain {W}all {QCD},
\newblock (2007), arXiv:0710.0536 [hep-lat].

\bibitem{Colangelo:2005gd}
G. Colangelo, S. D{\"u}rr and C. Haefeli, Finite volume effects for meson
  masses and decay constants,
\newblock Nucl. Phys. B721 (2005) 136, hep-lat/0503014.

\bibitem{Gasser:1987zq}
J. Gasser and H. Leutwyler, Spontaneously broken symmetries: {E}ffective
  {L}agrangians at {F}inite {V}olume,
\newblock Nucl. Phys. B307 (1988) 763.

\bibitem{Becirevic:2003wk}
D. Becirevic and G. Villadoro, Impact of the finite volume effects on the
  chiral behavior of {$f_K$} and {$B_K$},
\newblock Phys. Rev. D69 (2004) 054010, hep-lat/0311028.

\bibitem{Luscher:1985dn}
M. L{\"u}scher, Volume dependence of the energy spectrum in massive quantum
  field theories. 1. {S}table particle states,
\newblock Commun. Math. Phys. 104 (1986) 177.

\bibitem{Aubin:2003uc}
C. Aubin and C. Bernard, Pseudoscalar decay constants in staggered chiral
  perturbation theory,
\newblock Phys. Rev. D68 (2003) 074011, hep-lat/0306026.

\bibitem{Aubin:2003mg}
C. Aubin and C. Bernard, Pion and kaon masses in staggered chiral perturbation
  theory,
\newblock Phys. Rev. D68 (2003) 034014, hep-lat/0304014.

\bibitem{Follana:2006rc}
HPQCD, E. Follana et~al., Highly improved staggered quarks on the lattice, with
  applications to charm physics,
\newblock Phys. Rev. D75 (2007) 054502, hep-lat/0610092.

\bibitem{MILCconfigs}
C.W. Bernard et~al., The {QCD} spectrum with three quark flavors,
\newblock Phys. Rev. D64 (2001) 054506, hep-lat/0104002.

\bibitem{Sharpe:1995qp}
S.R. Sharpe and Y. Zhang, Quenched chiral perturbation theory for heavy-light
  mesons,
\newblock Phys. Rev. D53 (1996) 5125, hep-lat/9510037.

\bibitem{Leutwyler:1984je}
H. Leutwyler and M. Roos, Determination of the elements {$V_{us}$} and
  {$V_{ud}$} of the {K}obayashi-{M}askawa matrix,
\newblock Z. Phys. C25 (1984) 91.

\bibitem{Moulson:2007fs}
FlaviaNet Working Group on Kaon Decays, M. Moulson, {$V_{us}$} from kaon
  decays,
\newblock (2007), hep-ex/0703013.

\bibitem{Gasser:1984ux}
J. Gasser and H. Leutwyler, Low-energy expansion of meson form-factors,
\newblock Nucl. Phys. B250 (1985) 517.

\bibitem{Becirevic:2004ya}
D. Becirevic et~al., The {$K\to\pi$} vector form factor at zero momentum
  transfer on the lattice,
\newblock Nucl. Phys. B705 (2005) 339, hep-ph/0403217.

\bibitem{Becirevic:2005py}
D. Becirevic, G. Martinelli and G. Villadoro, The {A}demollo-{G}atto theorem
  for lattice semileptonic decays,
\newblock Phys. Lett. B633 (2006) 84, hep-lat/0508013.

\bibitem{Post:2001si}
P. Post and K. Schilcher, {$K_{l3}$} form factors at order {$p^6$} in chiral
  perturbation theory,
\newblock Eur. Phys. J. C25 (2002) 427, hep-ph/0112352.

\bibitem{Bijnens:2003uy}
J. Bijnens and P. Talavera, {$K_{l3}$} decays in chiral perturbation theory,
\newblock Nucl. Phys. B669 (2003) 341, hep-ph/0303103.

\bibitem{Cirigliano:2005xn}
V. Cirigliano et~al., The green function and {$SU(3)$} breaking in {$K_{l3}$}
  decays,
\newblock JHEP 04 (2005) 006, hep-ph/0503108.

\bibitem{Antonio:2007mh}
D.J. Antonio et~al., {$K_{l3}$} form factor with {$N_f=2+1$} dynamical domain
  wall fermions: {A} progress report,
\newblock (2007), hep-lat/0702026.

\bibitem{Juettner_Kaon}
RBC+UKQCD, A. J{\"u}ttner, {$K \to \pi$} semileptonic form factor with 2+1
  flavor domain wall fermions on the lattice,
\newblock PoS KAON (2007) 010.

\bibitem{James}
RBC+UKQCD, J. Zanotti, {$K\to\pi$} form factor with 2+1 dynamical domain wall
  fermions,
\newblock PoS LAT2007 (2007) 380.

\bibitem{Simula}
ETMC, S. Simula, Pseudo-scalar meson form factors with maximally twisted
  {W}ilson fermions at {$N_f = 2$},
\newblock (2007), arXiv:0710.0097 [hep-lat].

\bibitem{Morozov}
The QCDSF, D. Br{\"o}mmel et~al., Kaon semileptonic decay form factors from
  {$N_f = 2$} non- perturbatively {$O(a)$}-improved {W}ilson fermions,
\newblock (2007), arXiv:0710.2100 [hep-lat].

\bibitem{Tsutsui:2005cj}
JLQCD, N. Tsutsui et~al., Kaon semileptonic decay form factors in two-flavor
  {QCD},
\newblock PoS LAT2005 (2006) 357, hep-lat/0510068.

\bibitem{Dawson:2006qc}
C. Dawson et~al., Vector form factor in {$K_{l3}$} semileptonic decay with two
  flavors of dynamical domain-wall quarks,
\newblock Phys. Rev. D74 (2006) 114502, hep-ph/0607162.

\bibitem{Okamoto:2004df}
Fermilab Lattice, M. Okamoto, Full {CKM} matrix with lattice {QCD},
\newblock (2004), hep-lat/0412044.

\bibitem{Hill:2006bq}
R.J. Hill, Constraints on the form factors for {$K\to\pi l \nu$} and
  implications for {$|V_{us}|$},
\newblock Phys. Rev. D74 (2006) 096006, hep-ph/0607108.

\bibitem{Boyle:2007wg}
P.A. Boyle et~al., Hadronic form factors in lattice {QCD} at small and
  vanishing momentum transfer,
\newblock JHEP 05 (2007) 016, hep-lat/0703005.

\bibitem{Bedaque:2004kc}
P.F. Bedaque, Aharonov-bohm effect and nucleon nucleon phase shifts on the
  lattice,
\newblock Phys. Lett. B593 (2004) 82, nucl-th/0402051.

\bibitem{Sachrajda:2004mi}
C.T. Sachrajda and G. Villadoro, Twisted boundary conditions in lattice
  simulations,
\newblock Phys. Lett. B609 (2005) 73, hep-lat/0411033.

\bibitem{Bedaque:2004ax}
P.F. Bedaque and J.W. Chen, Twisted valence quarks and hadron interactions on
  the lattice,
\newblock Phys. Lett. B616 (2005) 208, hep-lat/0412023.

\bibitem{Flynn:2005in}
UKQCD, J.M. Flynn, A. J{\"u}ttner and C.T. Sachrajda, A numerical study of
  partially twisted boundary conditions,
\newblock Phys. Lett. B632 (2006) 313, hep-lat/0506016.

\bibitem{deDivitiis:2004kq}
G.M. de~Divitiis, R. Petronzio and N. Tantalo, On the discretization of
  physical momenta in lattice {QCD},
\newblock Phys. Lett. B595 (2004) 408, hep-lat/0405002.

\bibitem{Guadagnoli:2005be}
D. Guadagnoli, F. Mescia and S. Simula, Lattice study of semileptonic form
  factors with twisted boundary conditions,
\newblock Phys. Rev. D73 (2006) 114504, hep-lat/0512020.

\bibitem{Foster:1998vw}
UKQCD, M. Foster and C. Michael, Quark mass dependence of hadron masses from
  lattice {QCD},
\newblock Phys. Rev. D59 (1999) 074503, hep-lat/9810021.

\bibitem{McNeile:2006bz}
UKQCD, C. McNeile and C. Michael, Decay width of light quark hybrid meson from
  the lattice,
\newblock Phys. Rev. D73 (2006) 074506, hep-lat/0603007.

\bibitem{Jamin:2004re}
M. Jamin, J.A. Oller and A. Pich, Order {$p^6$} chiral couplings from the
  scalar {$K\to \pi$} form factor,
\newblock JHEP 02 (2004) 047, hep-ph/0401080.

\bibitem{Kl3-upcoming}
P.A. Boyle et~al., {$K_{l3}$} semileptonic form factor from 2+1 flavour lattice
  {QCD},
\newblock (2007), arXiv:0710.5136 [hep-lat].

\bibitem{Mescia:2007ku}
F. Mescia and f.t.F.W.G.o.K. Decays, Precision tests with {$K_{l3}$} and
  {$K_{l2}$} decays,
\newblock (2007), arXiv:0710.5620 [hep-ph].

\bibitem{FNCKM}
FlaviaNet Kaon Decay Working Group, http://www.lnf.infn.it/wg/vus/.

\bibitem{Fanti:1999nm}
NA48, V. Fanti et~al., A new measurement of direct {$CP$} violation in two pion
  decays of the neutral kaon,
\newblock Phys. Lett. B465 (1999) 335, hep-ex/9909022.

\bibitem{AlaviHarati:1999xp}
KTeV, A. Alavi-Harati et~al., Observation of direct {$CP$} violation in
  {$K_{S,L}\to\pi\pi$} decays,
\newblock Phys. Rev. Lett. 83 (1999) 22, hep-ex/9905060.

\bibitem{Christenson:1964fg}
J.H. Christenson et~al., Evidence for the {$2\pi$} decay of the {$K_2^0$}
  meson,
\newblock Phys. Rev. Lett. 13 (1964) 138.

\bibitem{Charles:2004jd}
CKMfitter Group, J. Charles et~al., {CP} violation and the {CKM} matrix:
  Assessing the impact of the asymmetric {B} factories,
\newblock Eur. Phys. J. C41 (2005) 1, hep-ph/0406184.

\bibitem{Martinelli:1983ac}
G. Martinelli, The four fermion operators of the weak hamiltonian on the
  lattice and in the continuum,
\newblock Phys. Lett. B141 (1984) 395.

\bibitem{Becirevic:2000cy}
D. Becirevic et~al., {$K^0-\bar K^0$} mixing with {W}ilson fermions without
  subtractions,
\newblock Phys. Lett. B487 (2000) 74, hep-lat/0005013.

\bibitem{Flynn:2004au}
UKQCD, J.M. Flynn, F. Mescia and A.S.B. Tariq, Sea quark effects in {$B_K$}
  from {$N_f=2$} clover-improved {W}ilson fermions,
\newblock JHEP 11 (2004) 049, hep-lat/0406013.

\bibitem{Mescia:2005ew}
F. Mescia et~al., Kaon {B}-parameter with {$N_f=2$} dynamical {W}ilson
  fermions,
\newblock PoS LAT2005 (2006) 365, hep-lat/0510096.

\bibitem{Frezzotti:2000nk}
ALPHA, R. Frezzotti et~al., Lattice {QCD} with a chirally twisted mass term,
\newblock JHEP 08 (2001) 058, hep-lat/0101001.

\bibitem{Dimopoulos:2006dm}
ALPHA, P. Dimopoulos et~al., A precise determination of {$B_K$} in quenched
  {QCD},
\newblock Nucl. Phys. B749 (2006) 69, hep-ph/0601002.

\bibitem{Dimopoulos:2007cn}
P. Dimopoulos et~al., Flavour symmetry restoration and kaon weak matrix
  elements in quenched twisted mass {QCD},
\newblock Nucl. Phys. B776 (2007) 258, hep-lat/0702017.

\bibitem{Guagnelli:2005zc}
ALPHA, M. Guagnelli et~al., Non-perturbative renormalization of left-left
  four-fermion operators in quenched lattice {QCD},
\newblock JHEP 03 (2006) 088, hep-lat/0505002.

\bibitem{VandeWater:2005uq}
R.S. Van~de Water and S.R. Sharpe, {$B_K$} in staggered chiral perturbation
  theory,
\newblock Phys. Rev. D73 (2006) 014003, hep-lat/0507012.

\bibitem{Gamiz:2006sq}
HPQCD, E. Gamiz et~al., Unquenched determination of the kaon parameter {$B_K$}
  from improved staggered fermions,
\newblock Phys. Rev. D73 (2006) 114502, hep-lat/0603023.

\bibitem{Kim:2006ck}
J. Kim et~al., {$B_K$} in unquenched {QCD} using improved staggered fermions,
\newblock PoS LAT2006 (2006) 086, hep-lat/0610057.

\bibitem{Hasenfratz:2001hp}
A. Hasenfratz and F. Knechtli, Flavor symmetry and the static potential with
  hypercubic blocking,
\newblock Phys. Rev. D64 (2001) 034504, hep-lat/0103029.

\bibitem{Lee:2003sk}
W.J. Lee and S.R. Sharpe, Perturbative matching of staggered four-fermion
  operators with hypercubic fat links,
\newblock Phys. Rev. D68 (2003) 054510, hep-lat/0306016.

\bibitem{Lee:2006cm}
W. Lee, Progress in kaon physics on the lattice,
\newblock PoS LAT2006 (2006) 015, hep-lat/0610058.

\bibitem{Aoki:2005ga}
Y. Aoki et~al., The kaon {B}-parameter from quenched domain-wall {QCD},
\newblock Phys. Rev. D73 (2006) 094507, hep-lat/0508011.

\bibitem{Antonio:2007pb}
RBC and UKQCD, D.J. Antonio et~al., Neutral kaon mixing from 2+1 flavor domain
  wall {QCD},
\newblock (2007), hep-ph/0702042.

\bibitem{Antonio}
RBC+UKQCD, D. Antonio, The kaon {B}-parameter from domain wall fermions,
\newblock PoS LAT2007 (2007) 344.

\bibitem{Cohen}
RBC+UKQCD, S. Cohen, {$B_K$} on 2+1-flavor {I}wasaki {DWF} lattices,
\newblock PoS LAT2007 (2007) 347.

\bibitem{Yamada}
JLQCD, N. Yamada et~al., {$B_K$} with dynamical overlap fermions,
\newblock (2007), arXiv:0710.0462 [hep-lat].

\bibitem{AliKhan:2001wr}
CP-PACS, A. Ali~Khan et~al., Kaon {B} parameter from quenched domain-wall
  {QCD},
\newblock Phys. Rev. D64 (2001) 114506, hep-lat/0105020.

\bibitem{Martinelli:1994ty}
G. Martinelli et~al., A general method for nonperturbative renormalization of
  lattice operators,
\newblock Nucl. Phys. B445 (1995) 81, hep-lat/9411010.

\bibitem{Aoki:2004ht}
Y. Aoki et~al., Lattice {QCD} with two dynamical flavors of domain wall
  fermions,
\newblock Phys. Rev. D72 (2005) 114505, hep-lat/0411006.

\bibitem{Aubin:2006hg}
C. Aubin, J. Laiho and R.S. Van~de Water, The kaon {B}-parameter in mixed
  action chiral perturbation theory,
\newblock Phys. Rev. D75 (2007) 034502, hep-lat/0609009.

\bibitem{Bar:2005tu}
O. B{$\"a$}r et~al., Chiral perturbation theory for staggered sea quarks and
  {G}insparg-{W}ilson valence quarks,
\newblock Phys. Rev. D72 (2005) 054502, hep-lat/0503009.

\bibitem{Aubin_firstBK}
C.A. Aubin, J. Laiho and R.S. Van~de Water, The kaon {$B$}-parameter from
  unquenched mixed action lattice {QCD},
\newblock (2007), arXiv:0710.1121 [hep-lat].

\bibitem{Pena}
F. Palombi et~al., Non-perturbative renormalization of static-light four-
  fermion operators in quenched lattice {QCD},
\newblock (2007), arXiv:0706.4153 [hep-lat].

\bibitem{Dimopoulos:2006es}
ALPHA, P. Dimopoulos et~al., Non-perturbative scale evolution of four-fermion
  operators in two-flavour {QCD},
\newblock PoS LAT2006 (2006) 158, hep-lat/0610077.

\bibitem{Dimopoulos:2006ma}
P. Dimopoulos et~al., Non-perturbative renormalisation of left-left
  four-fermion operators with {N}euberger fermions,
\newblock Phys. Lett. B641 (2006) 118, hep-lat/0607028.

\bibitem{Georgi}
H. Georgi,
\newblock Weak {I}nteractions and {M}odern {P}article {T}heory
  ({B}enjamin/{C}ummings, {M}enlo {P}ark, {C}alifornia, 1984) .

\bibitem{Buchalla:1995vs}
G. Buchalla, A.J. Buras and M.E. Lautenbacher, Weak decays beyond leading
  logarithms,
\newblock Rev. Mod. Phys. 68 (1996) 1125, hep-ph/9512380.

\bibitem{Giusti:2004an}
L. Giusti et~al., A strategy to study the role of the charm quark in explaining
  the {$\Delta I = 1/2$} rule,
\newblock JHEP 11 (2004) 016, hep-lat/0407007.

\bibitem{Hernandez:2006au}
P. Hernandez, Towards a quantitative understanding of the {$\Delta I = 1/2$}
  rule,
\newblock PoS LAT2006 (2006) 012, hep-lat/0610129.

\bibitem{Maiani:1990ca}
L. Maiani and M. Testa, Final state interactions from {E}uclidean correlation
  functions,
\newblock Phys. Lett. B245 (1990) 585.

\bibitem{Bernard:1985wf}
C.W. Bernard et~al., Application of chiral perturbation theory to {$K \to 2
  \pi$} decays,
\newblock Phys. Rev. D32 (1985) 2343.

\bibitem{Laiho:2002jq}
J. Laiho and A. Soni, On lattice extraction of {$K\to\pi\pi$} amplitudes to
  {$O(p^4)$} in chiral perturbation theory,
\newblock Phys. Rev. D65 (2002) 114020, hep-ph/0203106.

\bibitem{Lin:2002nq}
C.J.D. Lin et~al., {$K^+\to\pi^+\pi^0$} decays on finite volumes and at
  next-to-leading order in the chiral expansion,
\newblock Nucl. Phys. B650 (2003) 301, hep-lat/0208007.

\bibitem{Laiho:2003uy}
J. Laiho and A. Soni, Lattice extraction of {$K\to\pi\pi$} amplitudes to {NLO}
  in partially quenched and in full chiral perturbation theory,
\newblock Phys. Rev. D71 (2005) 014021, hep-lat/0306035.

\bibitem{Noaki:2001un}
CP-PACS, J.I. Noaki et~al., Calculation of non-leptonic kaon decay amplitudes
  from {$K\to\pi$} matrix elements in quenched domain-wall {QCD},
\newblock Phys. Rev. D68 (2003) 014501, hep-lat/0108013.

\bibitem{Blum:2001xb}
RBC, T. Blum et~al., Kaon matrix elements and {CP}-violation from quenched
  lattice {QCD}. {I}: {T}he 3-flavor case,
\newblock Phys. Rev. D68 (2003) 114506, hep-lat/0110075.

\bibitem{Bae:2005su}
T. Bae, J. Kim and W. Lee, Non-degenerate quark mass effect on {$B_K$} with a
  mixed action,
\newblock PoS LAT2005 (2006) 335, hep-lat/0510008.

\bibitem{Kim:2005st}
J. Kim, T. Bae and W. Lee, Calculating {$B_K$} using a mixed action,
\newblock PoS LAT2005 (2006) 338, hep-lat/0510007.

\bibitem{Boucaud:2004aa}
P. Boucaud et~al., An exploratory lattice study of {$\Delta I = 3/2\,$}
  {$K\to\pi\pi$} decays at next-to-leading order in the chiral expansion,
\newblock Nucl. Phys. B721 (2005) 175, hep-lat/0412029.

\bibitem{Noaki:2005zw}
J. Noaki, {$K\to\pi\pi$} from electroweak penguins in {$N_f=2$} domain- wall
  {QCD},
\newblock PoS LAT2005 (2006) 350, hep-lat/0510019.

\bibitem{Kim}
C. Kim, {$\Delta I=1/2$} {$K \rightarrow \pi\pi$} decays at next-to-leading
  order in chiral perturbation theory,
\newblock PoS LAT2007 (2007) 357, arXiv:0710.2519 [hep-lat].

\end{thebibliography}
\end{document}